\documentclass[11pt]{article}
\usepackage[utf8]{inputenc}

\usepackage{threeparttable}
\usepackage{amsxtra}
\usepackage{amsthm}
\usepackage{amssymb}
\usepackage{amsfonts}
\usepackage{graphicx}    
\usepackage{rotating}
\usepackage{color}
\usepackage{epsfig}
\usepackage{subfigure}   
\usepackage{verbatim}
\usepackage[numbers]{natbib}
\usepackage{algorithm,algcompatible}
\algnewcommand\INPUT{\item[\textbf{Input:}]}
\algnewcommand\OUTPUT{\item[\textbf{Output:}]}%
\usepackage{float}
\usepackage[printonlyused]{acronym} 
\usepackage{setspace}  
\usepackage{mathtools}
\usepackage{geometry}     
\usepackage{natbib}

\usepackage{indentfirst} 
\usepackage{color}
\usepackage[dvipsnames]{xcolor}
\usepackage{tikz}

\DeclareRobustCommand{\blackline}{\raisebox{2pt}{\tikz{\draw[-,black!40!Black,solid,line width = 0.9pt](0,0) -- (5mm,0);}}}
\DeclareRobustCommand{\blueline}{\raisebox{2pt}{\tikz{\draw[RoyalBlue,solid,line width = 0.9pt](0,0) -- (5mm,0);}}}
\DeclareRobustCommand{\orangeline}{\raisebox{2pt}{\tikz{\draw[BurntOrange,solid,line width = 0.9pt](0,0) -- (5mm,0);}}}
\DeclareRobustCommand{\greenline}{\raisebox{2pt}{\tikz{\draw[Green,solid,line width = 0.9pt](0,0) -- (5mm,0);}}}
\DeclareRobustCommand{\redline}{\raisebox{2pt}{\tikz{\draw[Red,solid,line width = 0.9pt](0,0) -- (5mm,0);}}}

\DeclareRobustCommand{\blackdashedline}{\raisebox{2pt}{\tikz{\draw[-,black!40!black,dashed,line width = 0.9pt](0,0) -- (5mm,0);}}}

\DeclareRobustCommand{\orangedashedline}{\raisebox{2pt}{\tikz{\draw[BurntOrange,dashed,line width = 0.9pt](0,0) -- (5mm,0);}}}
\DeclareRobustCommand{\greendashedline}{\raisebox{2pt}{\tikz{\draw[Green,dashed,line width = 0.9pt](0,0) -- (5mm,0);}}}
\DeclareRobustCommand{\reddashedline}{\raisebox{2pt}{\tikz{\draw[Red,dashed,line width = 0.9pt](0,0) -- (5mm,0);}}}
\DeclareRobustCommand{\purpledashedline}{\raisebox{2pt}{\tikz{\draw[Purple,dashed,line width = 0.9pt](0,0) -- (5mm,0);}}}
\DeclareRobustCommand{\browndashedline}{\raisebox{2pt}{\tikz{\draw[Brown,dashed,line width = 0.9pt](0,0) -- (5mm,0);}}}

\DeclareRobustCommand{\tikzcircle}[2][red,fill=red]{\tikz[baseline=-0.5ex]\draw[#1,radius=#2] (0,0) circle ;}

\usepackage{multirow}

\usepackage{graphicx}
\usepackage{caption}

\title{Multi-fidelity Bayesian experimental design to \\ quantify extreme-event statistics}
\author{Xianliang Gong, Zhou Zhang, Yulin Pan\footnote{yulinpan@umich.edu}}
\date{\small \textit{Department of Naval Architecture and Marine Engineering, University of Michigan, 48109, MI, USA}}

\begin{document}
\maketitle
\section*{Abstract}
In this work, we develop a multi-fidelity Bayesian experimental design framework to efficiently quantify the extreme-event statistics of an input-to-response (ItR) system with given input probability and expensive function evaluations. The key idea here is to leverage low-fidelity samples whose responses can be computed with a cost of a certain fraction of that for high-fidelity samples, in an optimized configuration to reduce the total computational cost. To accomplish this goal, we employ a multi-fidelity Gaussian process as the surrogate model of the ItR function, and develop a new acquisition based on which the optimized next sample can be selected in terms of its location in the sample space and the fidelity level. In addition, we develop an inexpensive analytical evaluation of the acquisition and its derivative, avoiding numerical integrations that are prohibitive for high-dimensional problems. The new method is tested in a bi-fidelity context for a series of synthetic problems with varying dimensions, low-fidelity model accuracy and computational costs. Comparing with the single-fidelity method and the bi-fidelity method with a pre-defined fidelity hierarchy, our method consistently shows the best (or among the best) performance for all the test cases. Finally, we demonstrate the superiority of our method in solving an engineering problem of estimating the extreme ship motion statistics in irregular waves, using computational fluid dynamics (CFD) with two different grid resolutions as the high and low fidelity models. 

\vspace{0.2cm}

\noindent \textit{Keywords}: extreme events, multi-fidelity models, Bayesian experimental design

\section{Introduction}
Extreme events are abnormally large system responses that can occur in many natural, engineering and other systems, with typical examples of rogue waves, ship capsize, extreme precipitation and structural failure. Although these events occur with a low probability, they may potentially result in catastrophic consequences to the environment, industry and society. Therefore, the quantification of the extreme-event statistics is of vital importance for assessment of the system reliability and engineering design to reduce the system failure probability \cite{farazmand2019extreme,ghil2011extreme,rahmstorf2011increase}.

Mathematically, with the system of interest characterized by a ``black-box'' input-to-response (ItR) function with known input probability, our goal is to efficiently evaluate the probability distribution of the system response especially regarding the extreme values. This is far from a trivial task mainly due to the expensive ItR function evaluation and the rareness of the extreme responses, resulting in a large number of required samples associated with potentially prohibitive computational cost. Many methods to overcome this difficulty (i.e., to reduce the number of required samples) have been developed in the context of sequential Bayesian experimental design (BED) 
\cite{chaloner1995bayesian}, or more broadly the active learning framework \cite{cohn1996active}. Two main components of these methods are  (1) a surrogate model, in many cases a Gaussian process \cite{rasmussen2003gaussian}, to approximate the ItR, and (2) a pre-defined acquisition function based on which the next-best sample is selected. Within this framework, early methods have been developed for reliability analysis (i.e., to compute the exceeding probability for system response above a given threshold), including AK-MCS \cite{echard2011ak}, EGRA \cite{bichon2008efficient} and many later improved variants \cite{hu2016global, wang2016gaussian,sun2017lif}. Recently, new acquisition functions \cite{mohamad2018sequential, gong2021full,sapsis2020output,blanchard2021bayesian} have also been developed which focus on obtaining the overall probability density function (PDF) of the response with an emphasis on the extreme-value portion (which is our purpose in this paper instead of the exceeding probability for a single threshold).  

Another consideration toward reducing the computational cost in the active learning framework is to leverage lower-fidelity models which calculate each response in a (small) fraction of cost (e.g., computational time or budget) of the high-fidelity model. Examples of such lower-fidelity models (as approximations to the high-fidelity counterparts) include (1) analytical models or numerical simulations as approximations to expensive physical experiments \cite{perdikaris2017nonlinear, marco2017virtual}; (2) coarse-grid computational fluid dynamics (CFD) simulations as approximations to fine-grid CFD simulations \cite{zheng2020multifidelity}; and (3) Reynolds-averaged Navier–Stokes models as approximations to large eddy simulations for turbulent flows \cite{wilcox1998turbulence}. Making use of the multi-fidelity Gaussian process \cite{kennedy2000predicting, perdikaris2017nonlinear} or neural network \cite{meng2020composite,meng2021multi} as surrogate models, multi-fidelity sampling algorithms have been developed for the purpose of global optimization \cite{song2019general, sarkar2019multifidelity}, function learning \cite{meng2021multi}, and contour detection \cite{marques2018contour}. In terms of the extreme-event statistics, the only related work within the multi-fidelity framework, to our knowledge, is \cite{yi2021active} which estimates the exceeding probability for reliability analysis. However, \cite{yi2021active} employs the sub-optimal acquisition function developed in AK-MCS which has been shown to be not only (much) less efficient than many improved algorithms later, but also not applicable to our purpose of obtaining the overall extreme-value portion of the response PDF. 

One of the key issues involved in the multi-fidelity sampling method is the determination of the fidelity level of each sample. Two types of methods have been considered regarding this issue, one to select the fidelity of next sample adaptively based on existing samples in order to reduce the overall computational cost (though in a heuristic manner) \cite{marques2018contour, marco2017virtual}, and the other to follow a pre-defined fidelity hierarchy (say pre-defined ratio and sequence of high and low fidelity samples in a bi-fidelity context) \cite{sarkar2019multifidelity, zheng2020multifidelity}. While the former type of method is developed in the hope of outperforming the latter one, there are not sufficient evidences to support the idea as systematic comparisons between the two types of methods are not available. It is one purpose of the current study, in the context of capturing the extreme-value response PDF, to systematically compare these two types of methods in determining the fidelity level of samples, along with the identification of their improvements relative to the single-fidelity algorithm \cite{blanchard2021output}.

Specifically, in the present work, we develop a multi-fidelity sequential BED framework for the quantification of the response PDF of an ItR system, with emphasis on the extreme-value portion. In particular, we use a multi-fidelity Gaussian process as the surrogate model, and develop an acquisition function (as a substantial extension to the single-fidelity function \cite{blanchard2021output}) which allows adaptive choice of both the location in the parameter space and fidelity level of the next sample. We also construct an analytical computation of the acquisition which avoids expensive numerical integration and enables high-dimensional implementation of the algorithm through gradient-based optimization. 
Our new method is systematically tested in a bi-fidelity context for a series of synthetic problems with varying dimensions, low-fidelity model accuracy and computational costs. 
We show that our bi-fidelity method outperforms the single-fidelity method in all test cases, and that our method for adaptive choice of fidelity level consistently performs among the best in all bi-fidelity runs with pre-defined fidelity hierarchy varying in a broad range. Finally, we demonstrate the coupling of our method with CFD to compute the PDF of extreme ship roll motion in irregular ocean waves. By using CFD simulations with two different grid resolutions as high and low fidelity models, we show that our bi-fidelity method achieves much faster convergence of the result (i.e., extreme response PDF) than the previous single-fidelity method. 

The python code for the algorithm, named MFGPextreme, is available on Github \footnote{https://github.com/umbrellagong/MFGPextreme}.  

\section{Method}
\subsection{Problem setup}
We consider a black-box ItR function $f(\mathbf{x}): \mathbb{R}^{d} \to \mathbb{R}$, with $\mathbf{x}\in \mathbb{R}^{d}$ representing the input parameters with known probability distribution $p_{\mathbf{x}}(\mathbf{x})$. We assume that we have a hierarchy of models $f_{1\sim s}(\mathbf{x})$ (from low to high fidelity) to compute $f(\mathbf{x})$, with $f_s(\mathbf{x})=f(\mathbf{x})$ and $f_i(\mathbf{x})$ having increasing deviations from $f(\mathbf{x})$ for $i=s-1, s-2, ..., 1$. In addition, the models $f_i(\mathbf{x})$ are associated with fixed computational costs $c_i$ which increases for $i=1, 2, ..., s$. The evaluation of these models are generally corrupted by noise, with observation $y_i$ defined as
\begin{equation}
    y_i = f_i(\mathbf{x}) + \epsilon_i, \;  \epsilon_i \sim \mathcal{N}(0, \gamma_i),  \quad \quad  i=1,2,\dots, s.
\label{noise}
\end{equation}
where $\epsilon_i$ is the Gaussian noise with variance $\gamma_i$ (representing observation error).  


Our quantity of interest is the PDF of the response $p_f(f)$, focusing on the tail part. Specifically, we aim to obtain an estimation $p_{f,est}(f)$ with minimized (see \cite{mohamad2018sequential})
\begin{equation}
    e = \int \Big|\log p_{f,est}(f) - \log p_{f,true}(f) \Big| df,
    \label{error_1}
\end{equation}
with $p_{f,true}(f)$ the true PDF of the response. We note that the $\log$ function in \eqref{error_1} acts on the ratio $p_{f,est}(f)/p_{f,true}(f)$, which is amplified when $p_{f,true}(f)$ is small, i.e., \eqref{error_1} emphasizes on the error in the small-probability portion (in many cases extreme-value portion) of the PDF.  

To compute $p_{f,est}(f)$, we can use a sequence of samples $y_i(\mathbf{x})$ with $i$ and $\mathbf{x}$ varying for each sample. Our objective is to find an optimized sequence, in terms of both $i$ and $\mathbf{x}$, such that $e$ is minimized under a given total computational cost $c$ (i.e., summation of computational cost $c_i$ over all members in the sequence). In general, there is no solutions to this type of problem that can be guaranteed to be global optimal, and the method we propose in this paper should be considered as a greedy algorithm that looks one step ahead of the existing samples. In particular, our method is based on  a multi-fidelity sequential BED, which involves two basic components: (1) an inexpensive surrogate model based on the multi-fidelity Gaussian process which infuses the information of multi-fidelity samples; (2) a new acquisitive function measuring the benefit (i.e., reduction in $e$) per computational cost, through the optimization of which the next-best sample can be selected in terms of both $i$ and $\mathbf{x}$. The two components are next described in detail in \S2.2 and \S2.3. In addition, in \S2.4, we develop an analytical formula to compute the acquisition function and its derivative with respect to $\mathbf{x}$, enabling the gradient-based optimization that is suitable for high-dimensional problems. 

\subsection{Surrogate model}
In this section, we briefly outline the multi-fidelity Gaussian process developed in \cite{kennedy2000predicting} as our surrogate model. Assume we have a dataset $\mathcal{D}=\{\mathcal{X}, \mathcal{Y} \}$ consisting of $s$ levels of model outputs $\mathcal{Y}=\{\mathcal{Y}_i\}_{i=1}^{s}$ at input positions $\mathcal{X}=\{\mathcal{X}_i\}_{i=1}^{s}$ sorted by increasing fidelity. 
The purpose of the multi-fidelity Gaussian process is to learn the underlying relation $f_i(\mathbf{x})$ from $\mathcal{D}$. This can be achieved through an auto-regressive scheme, which models $f_i(\mathbf{x})$ as
\begin{equation}
    f_i(\mathbf{x}) = \rho_{i-1} f_{i-1}(\mathbf{x}) + d_i(\mathbf{x}) \quad \quad  i=2,\dots,s, 
\label{AR}
\end{equation}
with $f_1(\mathbf{x}) \sim \mathcal{GP}(0, k_1(\mathbf{x},\mathbf{x}'))$ and $\{d_i(\mathbf{x}) \sim \mathcal{GP}(0, k_i(\mathbf{x},\mathbf{x}'))\}_{i=2}^{i=s}$ pairwise independent Gaussian processes, $\rho_{i-1}$ a scaling factor to quantity the correlation between $f_i$ and $f_{i-1}$. The kernels $k_i(\mathbf{x},\mathbf{x}')$ are defined as radial-basis functions
\begin{equation}
     k_i(\mathbf{x},\mathbf{x}') = \tau_i^2 \exp{\big(-\frac{1}{2}(\mathbf{x}- \mathbf{x}')^T \Lambda_i^{-1}(\mathbf{x}- \mathbf{x}')\big)},
\label{RBF}
\end{equation}
with $\tau_i$ and the diagonal matrix $\Lambda_i$ respectively representing the characteristic amplitude and length scales. $\{\tau_i, \Lambda_i, \gamma_i\}_{i=1}^{i=s}$ and $\{\rho_i\}_{i=1}^{i=s-1}$ are hyperparameters in the model and can be determined by maximizing the likelihood function $p(\mathbf{Y}=\mathcal{Y})$, where $\mathbf{Y}$ is a random vector of the (noise corrupted) surrogate with input at $\mathcal{X}$, satisfying a Gaussian distribution $\mathcal{N}(\mathbf{0}, {\rm{cov}}(\mathbf{Y}))$. Here we apply the shorthand notation ${\rm{cov}}(\mathbf{Y})={\rm{cov}}(\mathbf{Y},\mathbf{Y})$ to represent a covariance matrix for each pairwise random variables in $\mathbf{Y}$, which will be used throughout this paper.

The posterior prediction $f_i(\mathbf{x})$ given the dataset $\mathcal{D}$ can then be derived as a Gaussian process
\begin{equation}
    f_{i}(\mathbf{x})|\mathcal{D} \sim \mathcal{N}\big(\mathbb{E}(f_i(\mathbf{x})|\mathcal{D}), {\rm{cov}}(f_i(\mathbf{x}), f_i(\mathbf{x}')|\mathcal{D}) \big), \quad i=1,2,...,s
\label{mfgp}
\end{equation}
with analytically tractable mean and covariance (also defined across different fidelity levels)
\begin{align}
    \mathbb{E}(f_i(\mathbf{x})|\mathcal{D})  & = {\rm{cov}}(f_i(\mathbf{x}), \mathbf{Y}) {\rm{cov}}(\mathbf{Y})^{-1} \mathcal{Y},
\label{mean} \\  
    {\rm{cov}}\big(f_i(\mathbf{x}), f_j(\mathbf{x}') |\mathcal{D} \big) & = {\rm{cov}}\big(f_i(\mathbf{x}), f_j(\mathbf{x}')\big) - {\rm{cov}}\big(f_i(\mathbf{x}), \mathbf{Y} \big) {\rm{cov}}(\mathbf{Y})^{-1} {\rm{cov}}\big( \mathbf{Y}, f_j(\mathbf{x}') \big).
\label{cov}
\end{align}
In \eqref{mean} and \eqref{cov} (as well as the likelihood function), the covariances are computed as (or can be derived from):
\begin{equation}
     {\rm{cov}}(f_i(\mathbf{x}), f_j(\mathbf{x}')) 
    = \sum_{l=1}^{\min(i,j)} \pi_{ijl} k_l(\mathbf{x}, \mathbf{x}'),
\label{tf_cov_prior}
\end{equation}
where
\begin{equation}
\pi_{ijl}= \left\{
\begin{array}{ll}
      (\prod_{t=l}^{i-1} \rho_t) (\prod_{t=l}^{j-1} \rho_t) & l  \neq \min(i,j), \\
      \prod_{t=\min(i,j)}^{\max(i,j)-1} \rho_t &  l = \min(i,j), i \neq j, \\
      1 & l = \min(i,j), i=j. \\
\end{array} 
\right.
\end{equation}

We finally summarize the bi-fidelity counterpart of \eqref{mfgp} in Appendix A, which we will use in \S 3 for computation.

\subsection{Acquisition for sample selection}
Given the Gaussian process surrogate $f_{s}(\mathbf{x})|\mathcal{D}$ as in \eqref{mfgp} of the ItR, we can estimate the response PDF $p_{f|\mathcal{D}}(f)$. Our purpose is to select the next sample in terms of the fidelity level $i$ and the location $\tilde{\mathbf{x}}$ to significantly reduce the uncertainty in the extreme-value part of the response PDF (which is expected to lead to significantly smaller $e$ in \eqref{error_1}). In particular, this uncertainty can be estimated by (see previous work in the single-fidelity context \cite[][]{mohamad2018sequential})
\begin{equation}
    U(\mathcal{D}, i, \tilde{\mathbf{x}}) = \int |\log p_{f^{+}|\mathcal{D}, \overline{y}_i(\tilde{\mathbf{x}})}(f_{}) - \log p_{f^{-}|\mathcal{D}, \overline{y}_i(\tilde{\mathbf{x}})}(f_{})| \mathrm{d} f_{},
\label{acq_L}
\end{equation}
where, $\overline{y}_i = \mathbb{E}(f_i(\tilde{\mathbf{x}})|\mathcal{D})$ is the mean response computed by the surrogate $f_{i}(\mathbf{x})|\mathcal{D}$ from a hypothetical location $\tilde{\mathbf{x}}$ and fidelity $i$, $p_{f^{\pm}|\mathcal{D}, \overline{y}_i(\tilde{\mathbf{x}})}(f)$ are PDF bounds generated by upper and lower bounds (say two standard deviations away from the mean) of $f_{}|\mathcal{D}, \overline{y}_i(\tilde{\mathbf{x}})$.  

Using $U(\mathcal{D}, i, \tilde{\mathbf{x}})$ directly as the acquisition, however, involves significant computational cost (e.g., building a new Gaussian process $f_{}|\mathcal{D}, \overline{y}_i(\tilde{\mathbf{x}})$ for each hypothetical sample) even for single-fidelity problems. To address this issue, we extend the methodology developed for single-fidelity problems in \cite{sapsis2020output, blanchard2021output} to the multi-fidelity context. The first step is to introduce an upper bound as a proxy to $U$, defined as (proven in \cite{sapsis2020output} for single-fidelity applications)
\begin{equation}
    Q(\mathcal{D}, i, \tilde{\mathbf{x}}) = \int {\rm{var}}(f_{}(\mathbf{x})|\mathcal{D}, \overline{y}_i(\tilde{\mathbf{x}})) w(\mathbf{x}) \mathrm{d} \mathbf{x},
\label{Q}
\end{equation}
with
\begin{equation}
    w(\mathbf{x}) = \frac{p_{\mathbf{x}}(\mathbf{x})}{p_{\overline{f}}(\overline{f}(\mathbf{x}))},
\label{weight}
\end{equation}
where $\overline{f}(\mathbf{x}) = \mathbb{E}[f(\mathbf{x})|\mathcal{D}]$ represents the mean prediction \eqref{mean} with $i=s$. $Q$ measures the model uncertainty with emphasis on positions of large $w$, i.e. rare (usually large) response regions with significant input probability. We are interested in the reduction in $Q$ (i.e., the benefit) after adding the sample at $\tilde{\mathbf{x}}$ and $i$, formulated as 
\begin{align}
    B(i, \tilde{\mathbf{x}})  & = Q(\mathcal{D}) - Q(\mathcal{D}, i, \tilde{\mathbf{x}})
\nonumber \\
    & = \int \big({\rm{var}}(f_{}(\mathbf{x})|\mathcal{D})-{\rm{var}}(f_{}(\mathbf{x})|\mathcal{D}, \overline{y}_i(\tilde{\mathbf{x}})) \big) w(\mathbf{x}) \mathrm{d} \mathbf{x}
\nonumber \\
    & = \frac{1}{{\rm{var}}(y_i(\tilde{\mathbf{x}})|\mathcal{D})}\int {\rm{cov}}^2(f_{}(\mathbf{x}),f_i(\tilde{\mathbf{x}})|\mathcal{D}) w(\mathbf{x}) \mathrm{d} \mathbf{x}.
\label{tfgp_acq}
\end{align}
The derivation of the result in \eqref{tfgp_acq} is summarized in Appendix B, which makes use of the recursive update of the Gaussian process that is simpler than the derivation in \cite{blanchard2021output} for the single-fidelity method. We note that the expensive computations of the new posterior in \eqref{acq_L} and \eqref{Q} are not involved in \eqref{tfgp_acq}. Further reduction of computational cost by avoiding the numerical integration in \eqref{tfgp_acq} will be discussed shortly in \S2.4.

In general, one may expect that adding a high-fidelity sample is more beneficial than adding a low-fidelity sample at the same $\mathbf{x}$. While this is indeed generally true, we note that there exist some certain special situations in which adding a low-fidelity sample becomes more beneficial according to \eqref{tfgp_acq}. Such situations can occur when the function $d_i(\mathbf{x})$ in \eqref{AR} becomes uncorrelated, with a rigorous justification provided in Appendix C. To select the next best sample in terms of both location and fidelity level, we need an acquisition function taking into consideration both the benefit \eqref{tfgp_acq} and cost of the sample $c_i$. In particular, we solve an optimization problem
\begin{equation}
    \mathbf{x}^*, i^* = {\rm{argmax}}_{\tilde{\mathbf{x}} \in  \mathbb{R}^{d}, i \in \{1, 2, \dots s\}} \; B(i,\tilde{\mathbf{x}}) / c_i.
\label{opt}
\end{equation}

We remark that \eqref{opt} provides the optimal next sample in terms of the uncertainty reduction \eqref{tfgp_acq} per computational cost. Nevertheless, there is no guarantee that successively applying \eqref{opt} provides a global optimal solution although this type of fidelity-choice algorithm has been also applied for other purposes \cite{marques2018contour, marco2017virtual}. Therefore, the ultimate validity of \eqref{opt} needs to be tested in a sufficiently wide range of examples, especially against algorithms with a fixed fidelity hierarchy, which is one of our purposes in \S3.  

In solving \eqref{opt} as a combined discrete and continuous optimization problem, we first find the optimal location $\mathbf{x}$ for each fidelity $i$, i.e., $\mathbf{x}_i^*= {\rm{argmax}}_{\tilde{\mathbf{x}} \in  \mathbb{R}^{d}} \; B(i,\tilde{\mathbf{x}})$ for $i=1,2,\dots,s$, then we compare the benefit per cost for each fidelity level and find the optimal fidelity level, i.e., $\mathbf{x}^*, i^*= {\rm{argmax}}_{i \in \{1,2, \dots s\}} \; B(i,\mathbf{x}_i^*) / c_i$. 

While the solution procedure outlined above seems straightforward, there still exists a difficulty for applying the method to high-dimensional problems. The reason lies in that the numerical integration in \eqref{tfgp_acq} can become prohibitively expensive for high-dimensional $\mathbf{x}$. Furthermore, the high-dimensional optimization \eqref{opt} needs to rely on gradient-based algorithm where the derivative of \eqref{tfgp_acq} is also expensive to compute. To address these issues, analytical formulae for \eqref{tfgp_acq} and its derivative are much preferable, which will be discussed in the next section.

\subsection{Analytical formulae for \eqref{tfgp_acq} and its derivative}
To develop an analytical formula for \eqref{tfgp_acq}, we first substitute the expression of covariance function \eqref{cov} into  \eqref{tfgp_acq} and obtain
\begin{align}
    B(i,\tilde{\mathbf{x}}) & = 
    \frac{1}{{\rm{var}}(y_i(\tilde{\mathbf{x}})|\mathcal{D})} 
    \Big(
    \mathcal{K}(f_i(\tilde{\mathbf{x}}), f_i(\tilde{\mathbf{x}})) 
\nonumber \\
    & \quad + 
    {\rm{cov}}(f_i(\tilde{\mathbf{x}}), \mathbf{Y}) {\rm{cov}}(\mathbf{Y})^{-1} \big( \mathcal{K}(\mathbf{Y},\mathbf{Y})  {\rm{cov}}(\mathbf{Y})^{-1} {\rm{cov}}(\mathbf{Y}, f_i(\tilde{\mathbf{x}})) 
    - 2\mathcal{K}(\mathbf{Y}, f_i(\tilde{\mathbf{x}}))  \big )
    \Big),
\label{tfgp_acq2}
\end{align}
with
\begin{equation}
    \mathcal{K}(f_i(\mathbf{x}_1), f_j(\mathbf{x}_2)) = \int {\rm{cov}}\big(f_i(\mathbf{x}_1), f_{}(\mathbf{x})\big) {\rm{cov}}\big(f_{}(\mathbf{x}), f_j(\mathbf{x}_2))\big) w(\mathbf{x}) \mathrm{d} \mathbf{x}.
\label{K}
\end{equation}

We see that every term in \eqref{tfgp_acq2} is analytically tractable except the $\mathcal{K}$ function in \eqref{K} where the numerical integration is carried on. One idea to obtain an analytical form of $\mathcal{K}$, which has been suggested in the single-fidelity cases \cite{blanchard2021output,gong2021discussion}, is to approximate the $w(\mathbf{x})$ with a Gaussian mixture model \cite{Goodfellow-et-al-2016} with $n_{GMM}$ Gaussian functions:  
\begin{equation}
    w(\mathbf{x}) \approx \sum_{t=1}^{n_{GMM}} \alpha_t \mathcal{N}(\mathbf{x}; \mathbf{\mu}_t,\Sigma_t).
\label{gmm}
\end{equation}

This allows us to reformulate \eqref{K} as
\begin{equation}
     \mathcal{K}(f_i(\mathbf{x}_1), f_j(\mathbf{x}_2)) 
     \approx \sum_{t=1}^{n_{GMM}} \alpha_t G_t(f_i(\mathbf{x}_1), f_j(\mathbf{x}_2)),
\label{K_appro}
\end{equation}
with
\begin{equation}
     G_t(f_i(\mathbf{x}_1), f_j(\mathbf{x}_2)) = \int  {\rm{cov}}\big(f_i(\mathbf{x}_1), f(\mathbf{x})\big) {\rm{cov}}\big(f(\mathbf{x}), f_j(\mathbf{x}_2))\big) \mathcal{N}(\mathbf{x}; \mathbf{\mu}_t,\Sigma_t) \mathrm{d} \mathbf{x}.
\label{G}
\end{equation}

The problem now boils down to developing an analytical formula for \eqref{G}, which involves in the integrand the multiplication of two different multi-fidelity covariance functions and a Gaussian distribution function. This situation here is more complicated than that in the single-fidelity case \cite{blanchard2021output} where the problem is simplified by only involving two same single-fidelity covariance functions (e.g., analytical result from the latter case is already available \cite{mchutchon2013differentiating}). We summarize the detailed derivation of the analytical form of \eqref{G}, as well as the derivative $\partial B(i, \tilde{\mathbf{x}}) / \partial \tilde{\mathbf{x}}$ that can be derived in a similar manner, in Appendix D.

With analytical computation of $B(i, \tilde{\mathbf{x}})$ and $\partial B(i, \tilde{\mathbf{x}}) / \partial \tilde{\mathbf{x}}$ available, we can solve the optimization \eqref{opt} using gradient-based algorithm (which is more suitable for high-dimensional problems). In our current work, a gradient-based quasi-Newton method \cite{nocedal1980updating} with multiple starting points is used to solve \eqref{opt}, which completes the algorithm of the multi-fidelity BED method for extreme-event statistics. We finally summarize the full algorithm in Algorithm 1 and note that the algorithm reduces to a single-fidelity BED method for $s=1$. 

\begin{algorithm}
    \caption{Multi-fidelity BED for extreme-event statistics}
  \begin{algorithmic}
    \REQUIRE Number of initial samples $\{n_{init}(i)\}_{i=1}^{s}$, cost of each fidelity model $\{c_i\}_{i=1}^{s}$, total budget $c_{lim}$
    \INPUT Initial dataset $\mathcal{D}=\{\mathcal{X}, \mathcal{Y} \}$ with $\mathcal{X}=\{\mathcal{X}_i\}_{i=1}^{s}$ and $\mathcal{Y}=\{\mathcal{Y}_i\}_{i=1}^{s}$
    \STATE \textbf{Initialization} 
        $c_{total} = \sum_{i}  n_{init}(i) c_{i}$
    \WHILE{$c_{total} < c_{lim}$}
      \STATE 1. Train the surrogate model \eqref{AR} with $\mathcal{D}$ to obtain \eqref{mfgp}
      \STATE 2. Compute $w(\mathbf{x})$ in \eqref{weight} and approximate it with GMM model \eqref{gmm}
      \STATE 3. Solve the optimization \eqref{opt} to find the next-best sample $\{i^*, \boldsymbol{x}^*\}$
      \STATE 4. Evaluate the $i^*-$fidelity function to get $y_{i^*}(\mathbf{x}^*)$
      \STATE 5. Update the dataset $\mathcal{X}_{i^*} = \mathcal{X}_{i^*} \cup  \{\boldsymbol{x}^*\}$ and $ \mathcal{Y}_{i^*} = \mathcal{Y}_{i^*} \cup \{y_{i^*}(\mathbf{x}^*)\}$
      \STATE 6. $c_{total} = c_{total} + c_{i^*} $
    \ENDWHILE
\OUTPUT Compute the response PDF based on the surrogate model \eqref{mean} 
  \end{algorithmic}
\label{al}
\end{algorithm}

\section{Results}

In this section, we test our developed method in the context of bi-fidelity problems, i.e., $s=2$ and we use $f_1(\mathbf{x})=f_l(\mathbf{x})$, $c_1=c_l$ and $f_2(\mathbf{x})=f_h(\mathbf{x})$, $c_2=c_h$ for clarity. The tests are conducted for three synthetic problems with dimensions $d=1, 2, 8$ (\S3.1, \S3.2, \S3.3) and an engineering problem (of $d=2$) to evaluate extreme ship motion statistics in irregular waves with CFD of low and high resolutions as $f_l(\mathbf{x})$ and $f_h(\mathbf{x})$ (\S3.4). In synthetic problems, the true solution $p_{f,true}$ involved in \eqref{error_1} is obtained from a computation using a sufficiently large number of high-fidelity samples (since we assume $f(\mathbf{x})=f_h(\mathbf{x})$ discussed in \S2.1). In all cases, we compare the results from our method (bi-fidelity optimal sampling in both location and fidelity level as in Algorithm 1, hereafter termed ``BF-O'') to those from single-fidelity optimal sampling \cite{blanchard2021output} (Algorithm 1 with $s=1$, hereafter termed ``SF''). In addition, for the synthetic problems with $d=2$ and $8$, we further include the results from the bi-fidelity model with fixed ratio $n$ (as well as sequence) of low and high fidelity samples with locations optimized (Algorithm 1 but with optimization $\mathbf{x}_i^*= {\rm{argmax}}_{\tilde{\mathbf{x}} \in  \mathbb{R}^{d}} \; B(i,\tilde{\mathbf{x}})$ solved for fixed $i$ on each sample, hereafter termed ``BF-F$n$''). In these cases, we also vary the number $n$, the ratio $c_h/c_l$ and the low-fidelity accuracy level in a broad range, to assess the performance of our method BF-O in various situations for cases of intermediate to relatively high complexity.  

For all the methods in comparison, an initial data set is required to start the sequential BED procedure. To create a fair situation for comparison, we keep the cost to generate the initial dataset the same for all methods. In particular, the initial set is generated by a space-filling Latin Hypercube sampling method \cite{mckay2000comparison}, with $4d$ high-fidelity samples for SF method, and $2d$ high-fidelity and $2d \; c_h/c_l$ low-fidelity samples for BF-F$n$ and BF-O methods. Since the initial sample locations are not fixed, we will present the results from SF, BF-O and BF-F$n$ methods in terms of the average over 100 implementations with different initial datasets for the synthetic cases in \S3.1, \S3.2, and \S3.3.

\subsection{One-dimensional Forrester function}
\begin{figure}
\centering
\begin{minipage}[b]{0.34\linewidth}
\includegraphics[width = \linewidth]{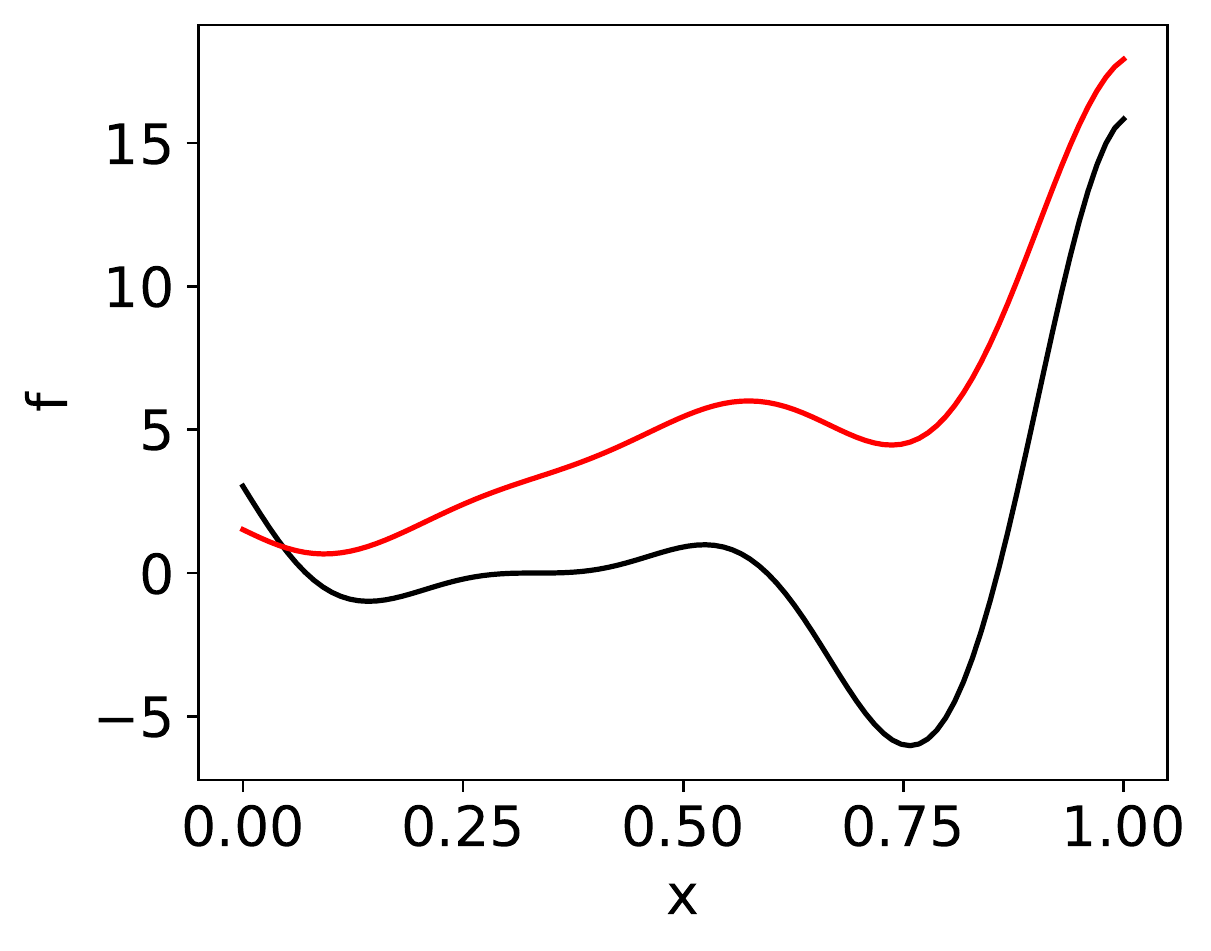}
\centering{(a)}
\end{minipage}
\begin{minipage}[b]{0.34\linewidth}
\includegraphics[width = \linewidth]{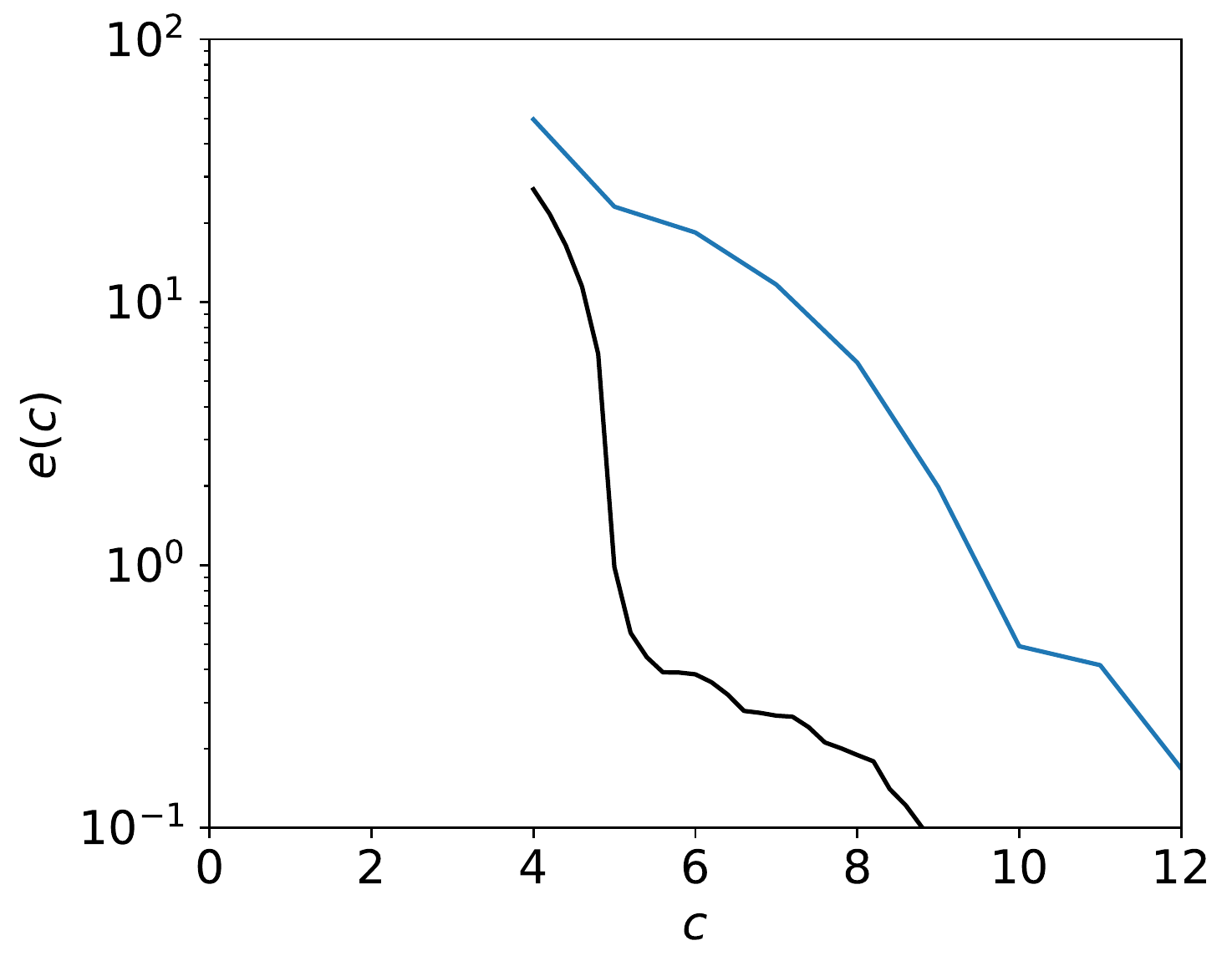}
\centering{(b)}
\end{minipage}
\begin{minipage}[b]{0.29\linewidth}
\includegraphics[width = \linewidth]{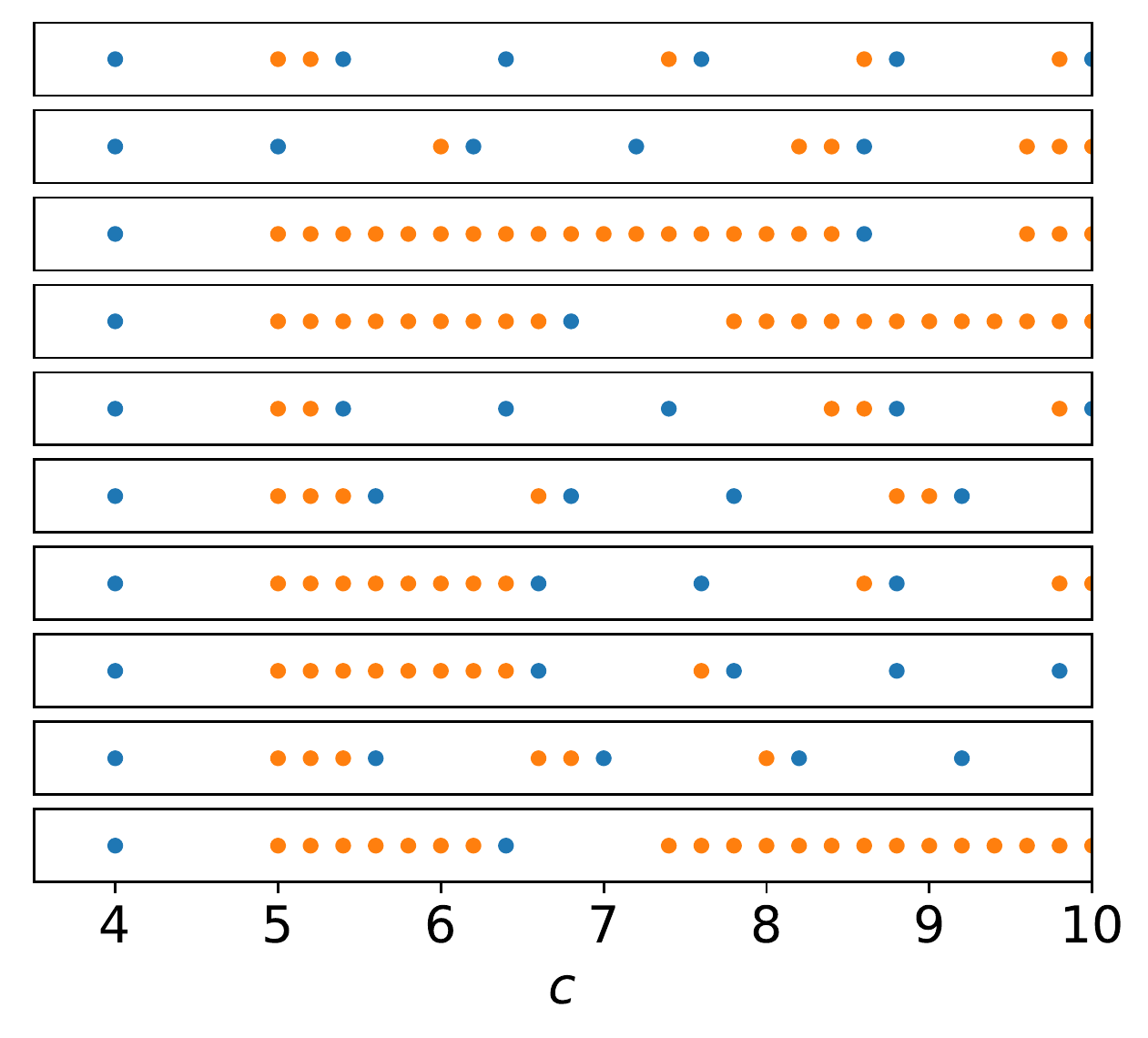}
\centering{(c)}
\end{minipage}
\caption{(a) The low fidelity function $f_l(x)$ (\redline) and high-fidelity function $f_h(x)$ (\blackline) in the 1D problem. (b) The corresponding error $e(c)$ computed by BF-O(\blackline) and SF(\blueline). (c) The sequence of high-fidelity (\tikzcircle{1pt, NavyBlue}) and low-fidelity (\tikzcircle{1pt, orange}) samples in ten experiments of BF-O.}
\label{fig:forrester_linear}
\end{figure}

We start the method validation from a one-dimensional (1D) Forrester function $f(\mathbf{x})$ that has been previously used to demonstrate the multi-fidelity global optimization \cite{forrester2007multi}. The high-fidelity ($f_h(\mathbf{x})=f(\mathbf{x})$, $c_h=1$) and low-fidelity ($f_l(\mathbf{x})$, $c_l=0.2$) models are constructed as (see figure \ref{fig:forrester_linear}(a)) 
\begin{align}
    f_h(x) & = (6x - 2)^2 \sin(12x - 4), \\
    f_l(x) & = 0.5 f_h(x) + 10x,
\label{forrester}
\end{align}
where the input $x$ is assumed to follow a Gaussian distribution with $p_{x}(x) = \mathcal{N}(0.5, 0.1)$. The results of BF-O and SF methods are shown in figure \ref{fig:forrester_linear}(b) in terms of the error $e$ (as in \eqref{error_1}) as a function of the total cost $c$ (i.e., summation of $c_l$ and $c_h$ for all samples). The BF-O method clearly outperforms the SF method to a large extent. For example, at $c=6$, the BF-O method achieves a value of $e$ that is nearly two orders of magnitude smaller than the SF method. In figure \ref{fig:forrester_linear}(c), we show 10 examples of the sequence of fidelity levels in samples by BF-O. While the sequences vary for different initial datasets, the BF-O algorithm exclusively selects a high-fidelity sample as the first sequential sample. This is consistent with an intuitive understanding that the algorithm tends to use three high-fidelity samples (2 in the initial dataset and 1 as selected) to learn the linear difference term $10x$ in \eqref{forrester} using a Gaussian process, which results in a significant reduction in $e$ at $c\approx 5$ in figure \ref{fig:forrester_linear}(b). 

While this simple example demonstrates the advantage of using BF-O method, it is only for a single case with fixed accuracy level of the low-fidelity model (in terms of \eqref{forrester}) and the cost ratio $c_h/c_l$. We will next use a two-dimensional (2D) case to test a much broader range of situations in the next section.

\subsection{Two-dimensional stochastic oscillator}
We consider a 2D function constructed from the solution of a stochastic oscillator equation, which has been previously used for testing the single-fidelity BED method (i.e., SF method) in  \cite{mohamad2018sequential, blanchard2021output}. In particular, the oscillator equation is formulated as
\begin{equation}
    \ddot{u} + \delta \dot{u} + F(u) = \xi(t),
\end{equation}
where $u(t)$ is the state variable, $F$ is a nonlinear restoring force defined by:
$$
F(u) = \left\{
\begin{aligned}
    & \alpha u           &
    &  if \; 0 \leq |u| \leq u_1 \\
    & \alpha u_1         &
    &  if \; u_1 \leq |u| \leq u_2  \\
    & \alpha u_1 + \beta(u-u2)^3 &
    &  if \; u_2 \leq|u|           \\
\end{aligned}
\right..
$$
The stochastic process $\xi(t)$, with a correlation function $\sigma_{\xi}^2 e^{-\tau^2/(2 l_{\xi}^2)}$, is approximated by a two-term Karhunen-Loeve expansion
\begin{equation}
    \xi(t) = \sum_{i=1}^{2} x_i \lambda_i \phi(t),
\end{equation}
with $\lambda_i$ and $\phi(t)$ respectively the eigenvalue and eigenfunction of the correlation function, $\mathbf{x}\equiv (x_1, x_2)$ is a standard normal variable as the input to the system, satisfying $p_{\mathbf{x}}(\mathbf{x}) = \mathcal{N}(\mathbf{0}, \mathrm{I})$ with $\mathrm{I}$ being a $2 \times 2$ identity matrix. The values of the parameters are kept the same as those in the single-fidelity work \cite{blanchard2021output} \footnote[1]{$\delta$=1.5, $\alpha$=1, $\beta$=0.1, $u_1$=0.5, $u_2$=1.5, $\sigma_{\xi}^2$=0.1, $l_\xi$=4}.

\begin{figure}
\centering
\begin{minipage}[b]{0.47\linewidth}
\includegraphics[width = \linewidth]{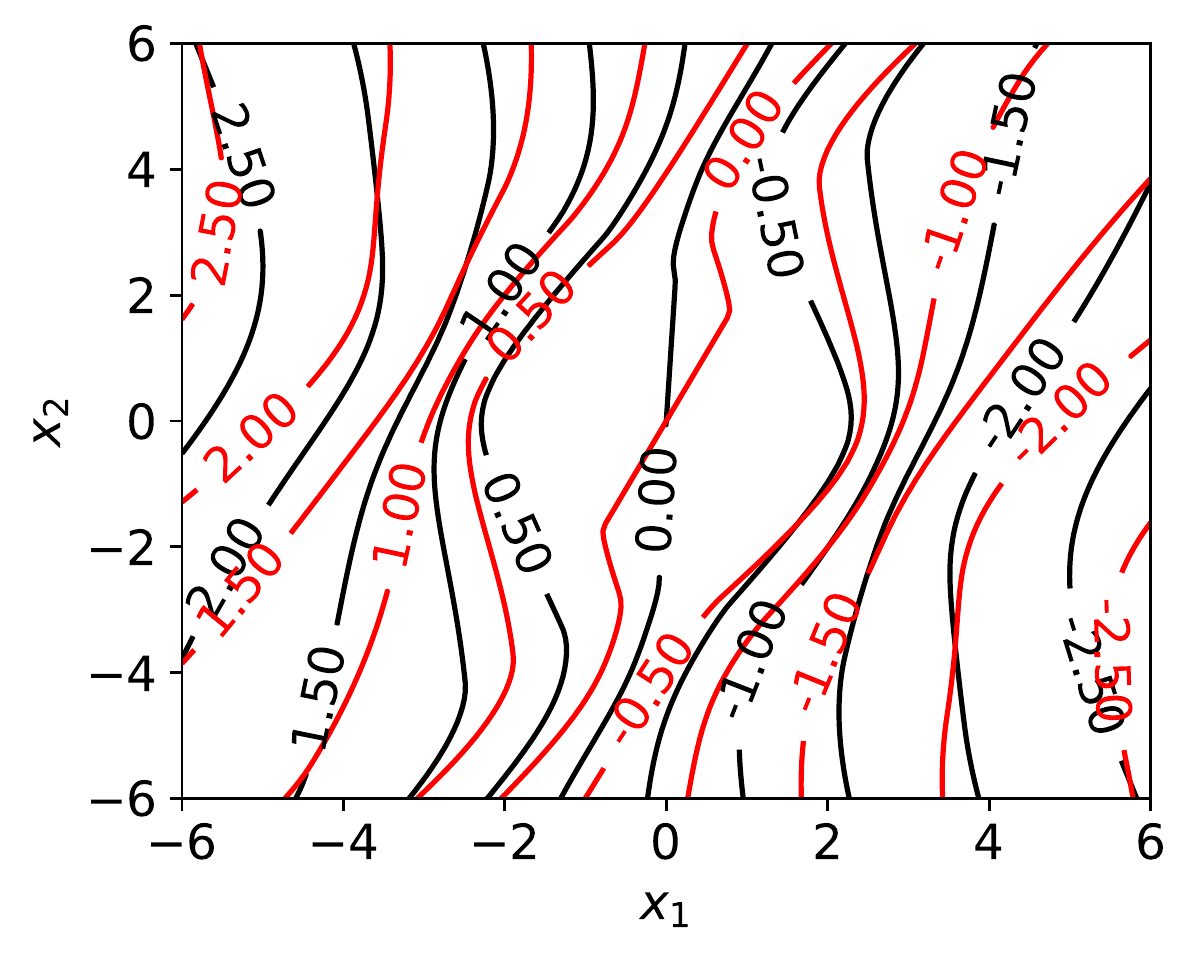}
\centering{(a)}
\end{minipage}
\begin{minipage}[b]{0.49\linewidth}
\includegraphics[width = \linewidth]{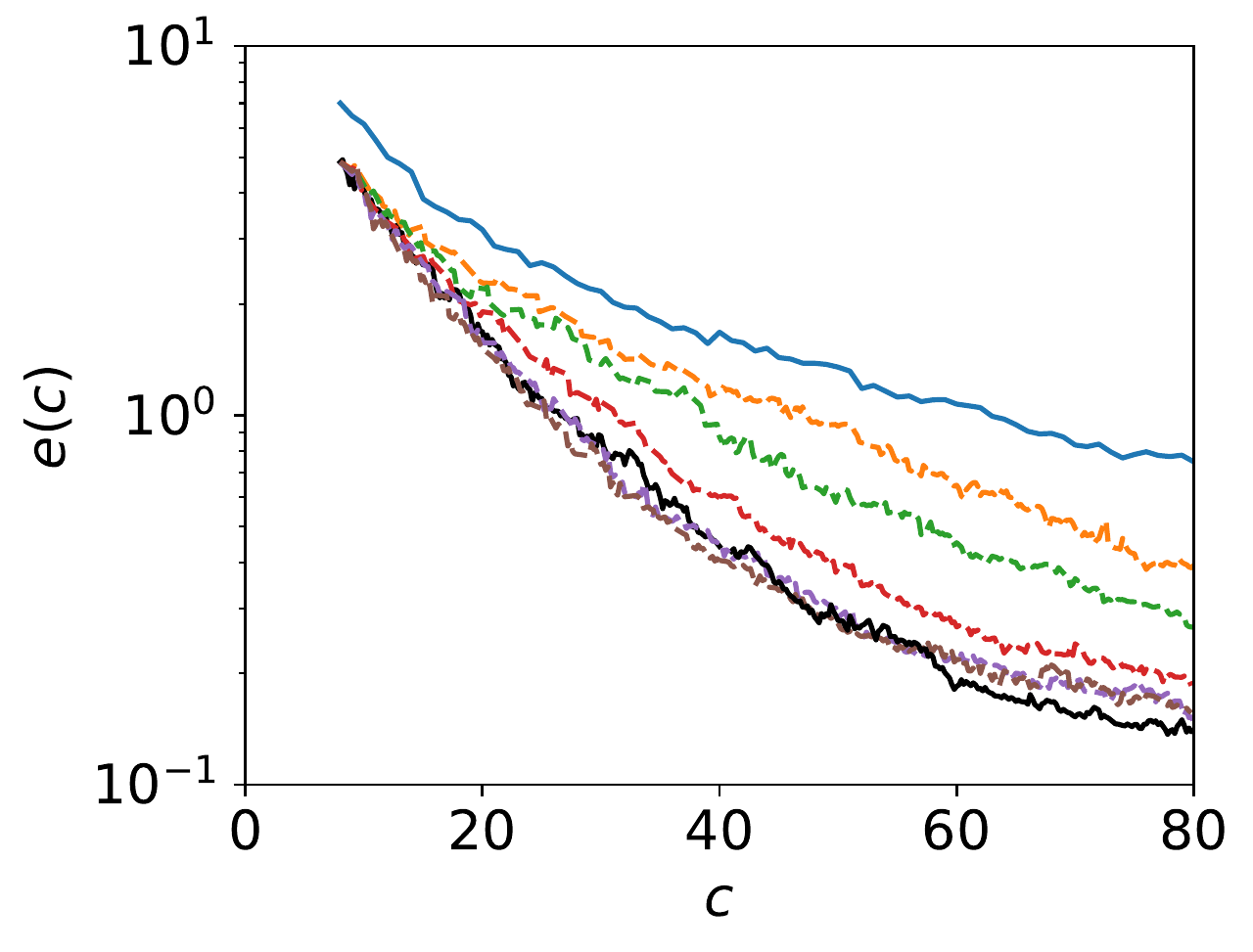}
\centering{(b)}
\end{minipage}
\begin{minipage}[b]{\linewidth}
\centering
\quad \quad
\includegraphics[width = 0.9\linewidth]{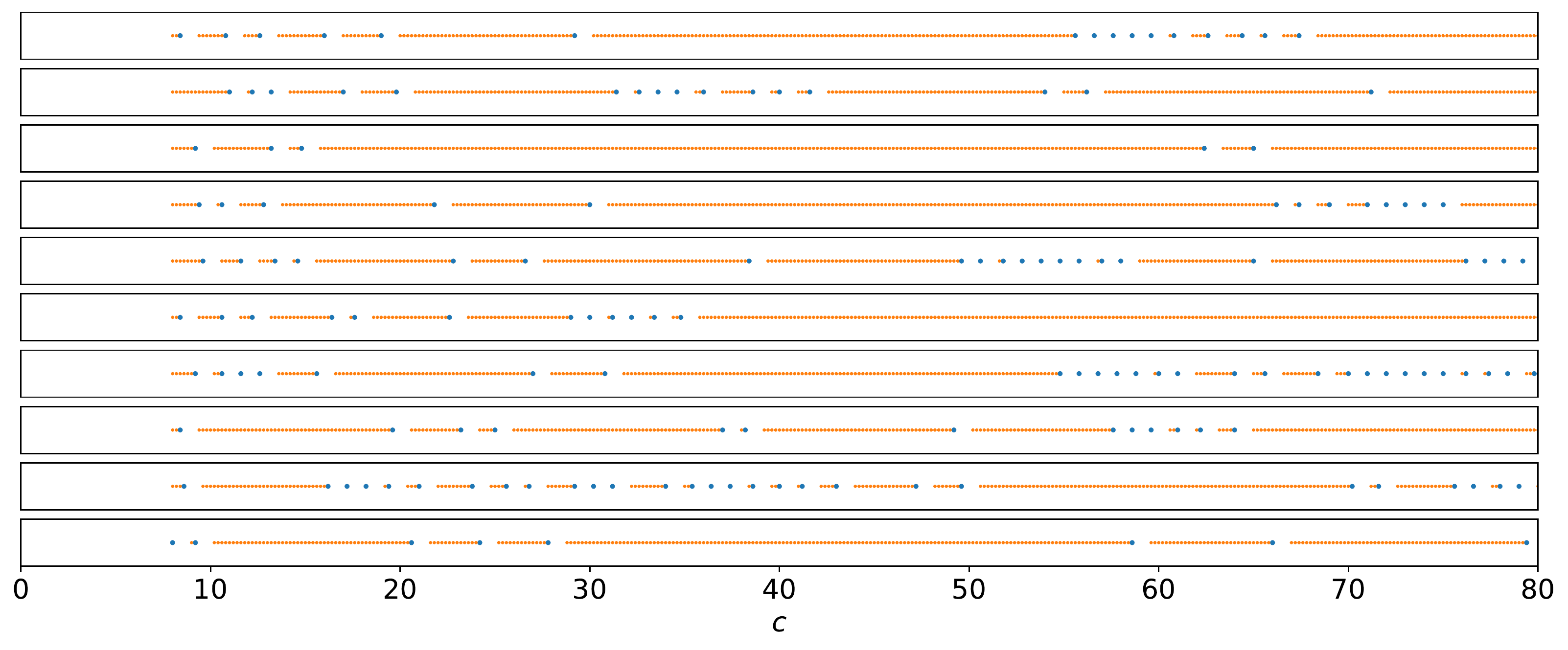}
\\ \centering{(c)}
\end{minipage}
\caption{(a) The low-fidelity (\redline) and high-fidelity (\blackline) functions with a linear difference in the 2D oscillator problem. (b) The corresponding error $e(c)$ computed by BF-O(\blackline), BF-F1(\orangedashedline), BF-F2(\greendashedline), BF-F5(\reddashedline), BF-F10(\purpledashedline), BF-F15 (\browndashedline) and SF(\blueline). (c) The sequence of high-fidelity (\tikzcircle{1pt, NavyBlue}) and low-fidelity (\tikzcircle{1pt, orange}) samples in ten experiments of BF-O.}
\label{fig:os_linear}
\end{figure}

The response of the system is considered as the mean value of $u(t;\mathbf{x})$ in the interval $[0,25]$, which serves as our high-fidelity model:
\begin{equation}
    f_h(\mathbf{x}) = \frac{1}{25} \int_{0}^{25}u(t;\mathbf{x}) \mathrm{d} t.
\label{2d_h}
\end{equation}
For our low-fidelity model, we construct a function $f_l(\mathbf{x})$ (to be varied later in this section) with a difference $d(\mathbf{x})$ from \eqref{2d_h}:
\begin{equation}
    f_l(\mathbf{x}) = \rho f_h(\mathbf{x}) + d(\mathbf{x}),
\label{2d_low}
\end{equation}
with $\rho=1$ and $d(\mathbf{x})$ chosen as a linear function $0.05 (x_1 + x_2)$ in this case. Both the $f_h(\mathbf{x})$ and $f_l(\mathbf{x})$ functions are shown in figure \ref{fig:os_linear}(a) to illustrate the functional forms and their difference. In this case, we use $c_l=0.2$ and $c_h=1$ as the computational cost of low and high fidelity models.

The error $e(c)$ is plotted in \ref{fig:os_linear}(b) for SF, BF-O and BF-F$n$ with $n$ varying from 1 to 15. We see that all bi-fidelity methods (BF-O and BF-F$n$) achieve acceleration on the error reduction (to different extents) compared to the SF method. For the BF-F$n$ method, faster convergence is observed for larger $n$ in the test range of $n\in[1, 15]$, but with much less benefit for $n$ increasing from 10 to 15. The BF-O method provides the best result, in terms of the error $e$ at cost $c=80$, although the BF-O result is somewhat less accurate than the BF-F$15$ result for smaller $c$ in the range of [25,50]. Accounting for all the sequence of fidelity levels (with 10 examples shown in \ref{fig:os_linear}(c)), the average ratio of high and low fidelity samples selected by BF-O is approximately 19.06, close to the value of $n=15$ which is found to be the best in BF-F$n$ in the test range.

\begin{table}
\begin{center}
\caption{Setting of cases in the 2D oscillator problem.}
\begin{threeparttable}[t]
\begin{tabular}{|c|c|c|c|c|c|c|c|}
\hline
no. &  $d(\mathbf{x})$ & $\rho$                  & $\delta$                     & $c_h/c_l$  & \begin{tabular}[c]{@{}c@{}}rank of BF-O\\ at cost 80\end{tabular}
& \begin{tabular}[c]{@{}c@{}}$n_l / n_h$\\ in BF-O\end{tabular}
&  \begin{tabular}[c]{@{}c@{}} best $n$\\ in BF-F$n$\end{tabular}\\ \hline
1        & $\delta (x_1 + x_2)$                         & \multirow{9}{*}{1} & 0.05                                                & \multirow{3}{*}{5} & 1 & 19.06 & 15\\ \cline{1-2} \cline{4-4} \cline{6-8} 
2        & \multirow{9}{*}{$\delta \sin (x_1 + x_2)$} &                    & 0.02                                                &                    & 2 & 11.95 & 5\\ \cline{1-1} \cline{4-4} \cline{6-8} 
3        &                                            &                    & 0.05                                                &                    & 1 & 7.19 & 5 \\ \cline{1-1} \cline{4-8} 
4        &                                            &                    & \multirow{4}{*}{0.1}                                & 2                   & 1 & 0.47 & 1\\ \cline{1-1} \cline{5-8} 
5        &                                            &                    &                                                     & 5                   & 1 & 4.74 & 5\\ \cline{1-1} \cline{5-8} 
6        &                                            &                    &                                                     & 8                   & 1 & 5.93 & 5\\ \cline{1-1} \cline{5-8} 
7        &                                            &                    &                                                     & 10                  & 1 & 6.85 & 10\\ \cline{1-1} \cline{4-8} 
8        &                                            &                    & 0.2                                                 & \multirow{3}{*}{5}  & 2 & 3.19 & 5\\ \cline{1-1} \cline{4-4} \cline{6-8} 
9        &                                            &                    & 0.4                                                 &                      & 2  & 2.16 & 5\\ \cline{1-1} \cline{3-4} \cline{6-8} 
10       &                                            & 1.5                & 0.1                                                &                     & 1  & 6.91 & 5\\ \hline
\end{tabular}
\end{threeparttable}
\label{table:2d}
\end{center}
\end{table}

\begin{figure}
\centering
\begin{minipage}[b]{0.45\linewidth}
\includegraphics[width = \linewidth]{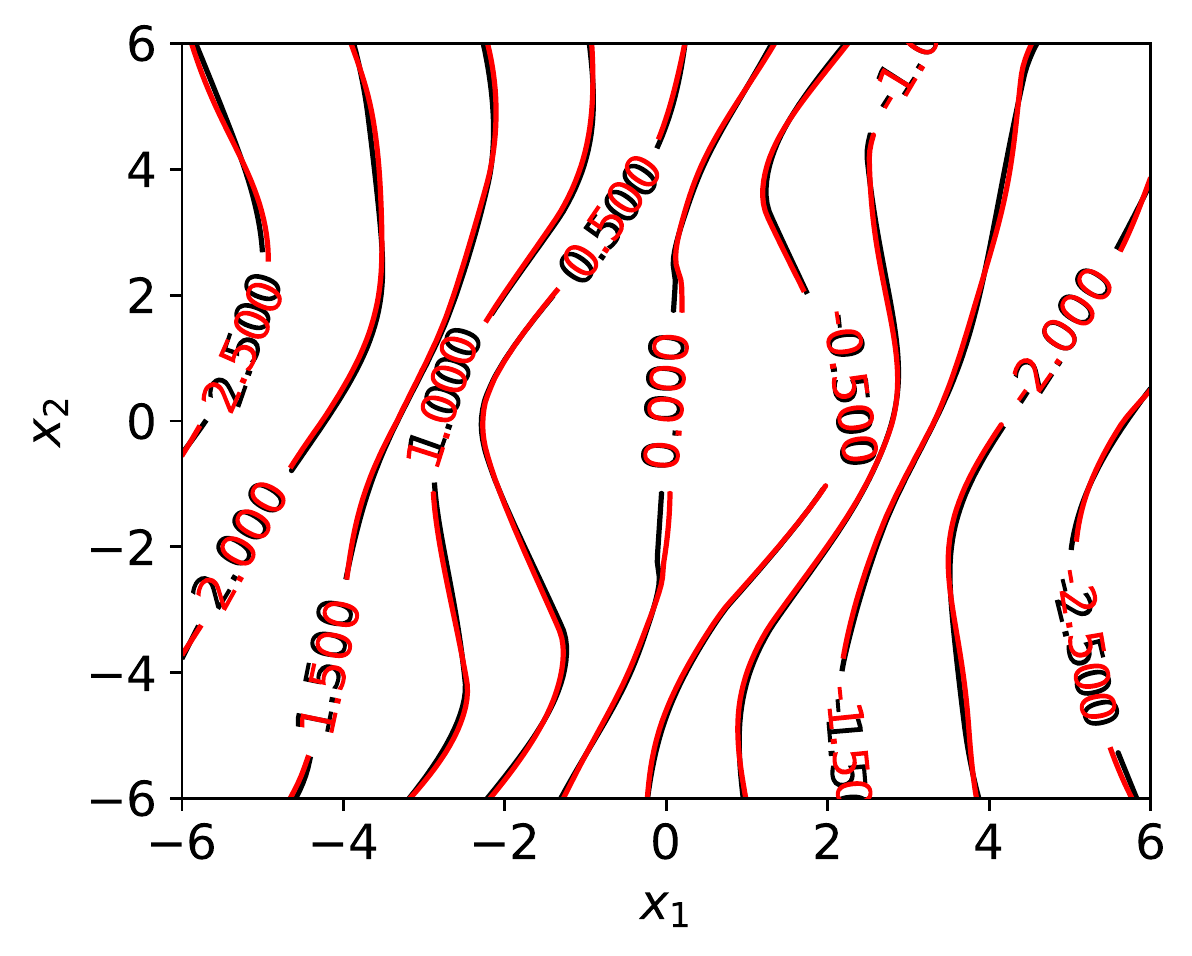}
\end{minipage}
\centering{\quad (a)}
\begin{minipage}[b]{0.47\linewidth}
\includegraphics[width = \linewidth]{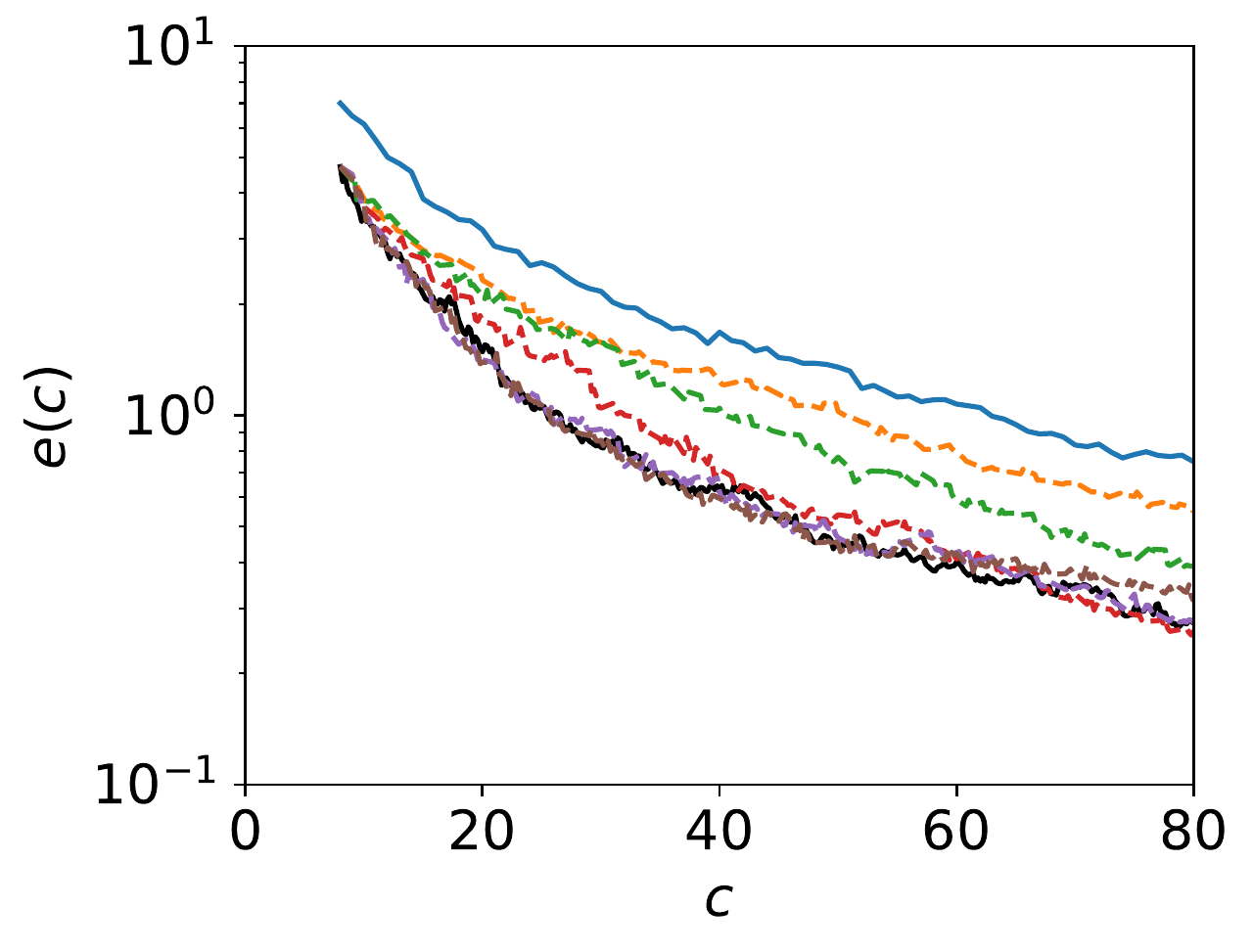}
\end{minipage}
\begin{minipage}[b]{0.45\linewidth}
\includegraphics[width = \linewidth]{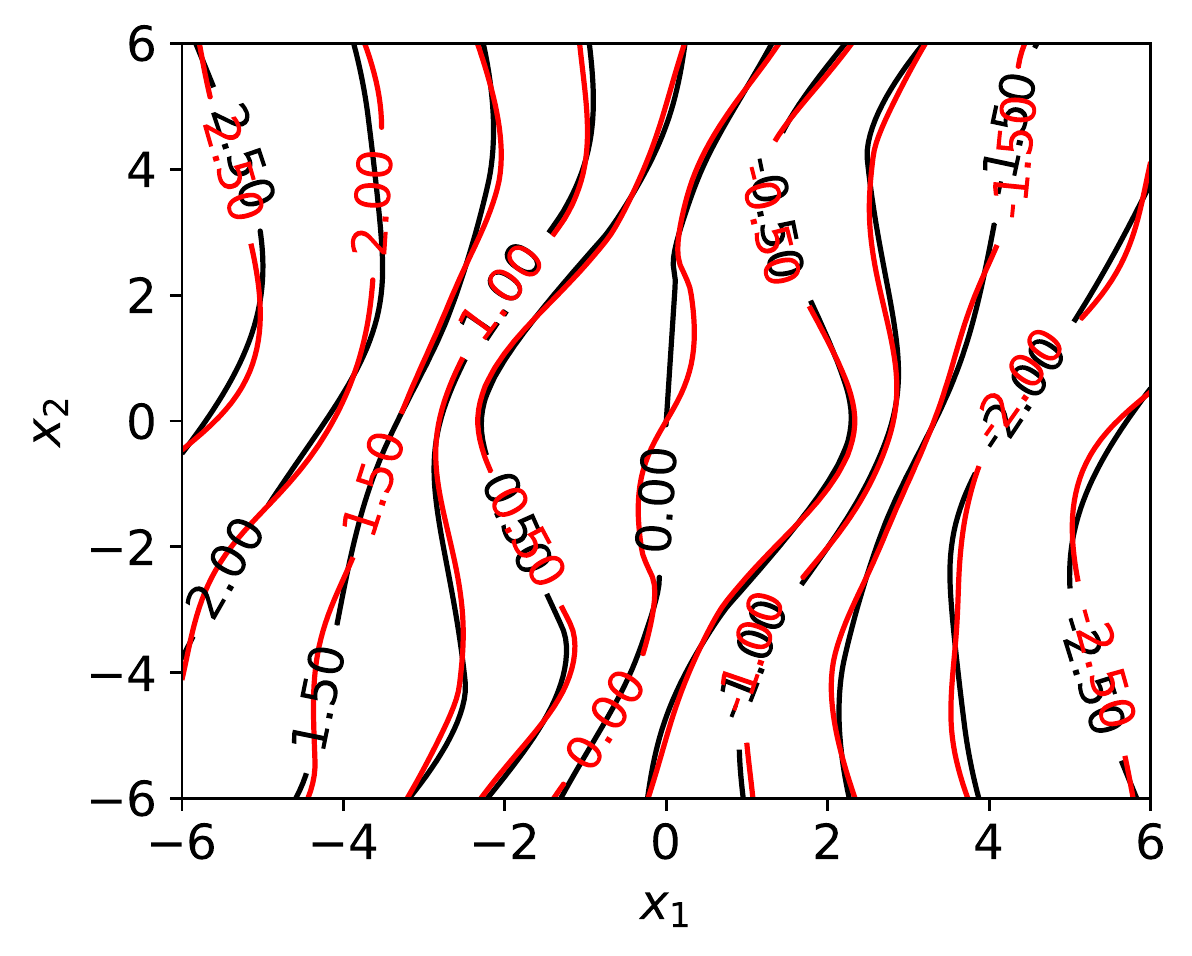}
\end{minipage}
\centering{\quad (b)}
\begin{minipage}[b]{0.47\linewidth}
\includegraphics[width = \linewidth]{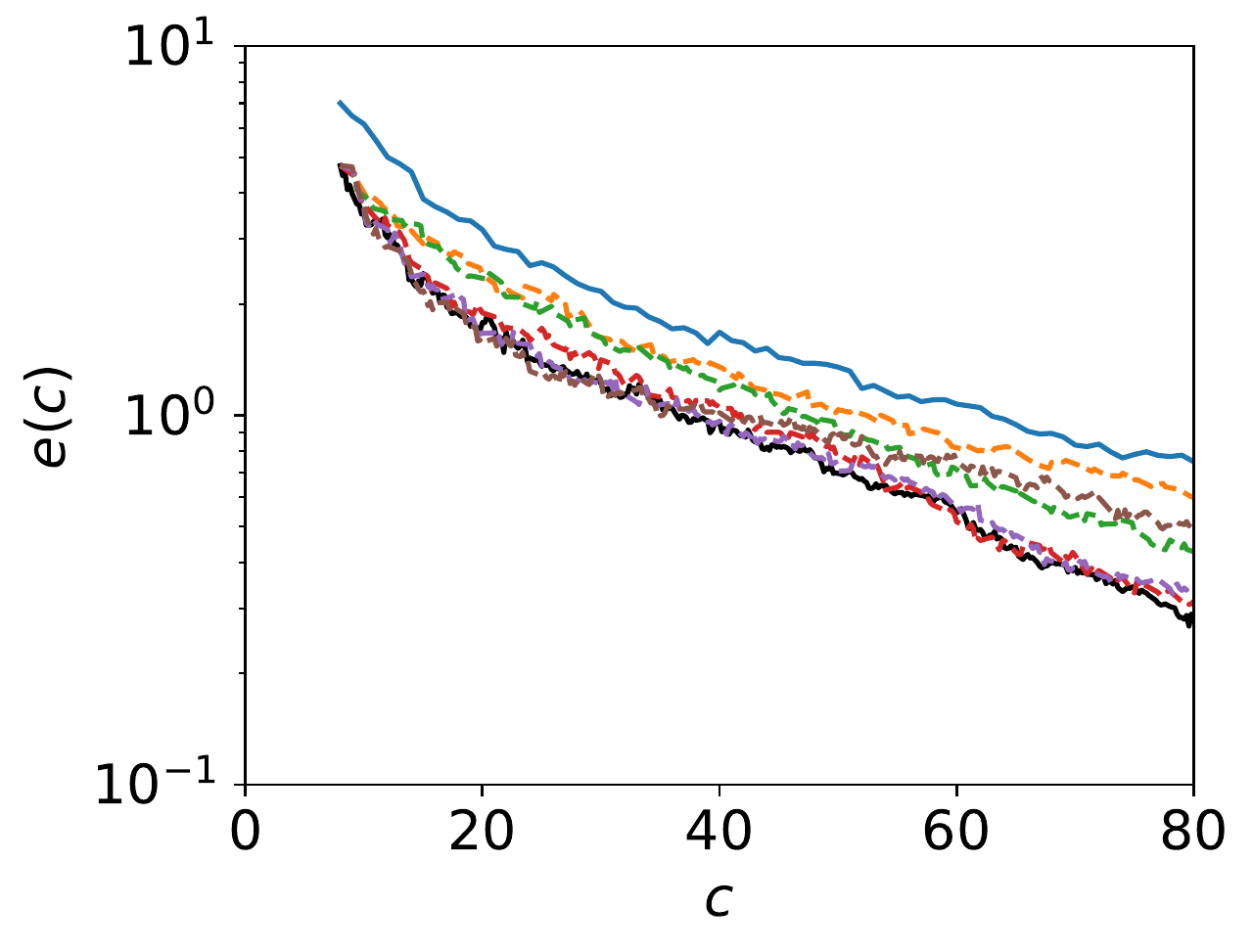}
\end{minipage}
\begin{minipage}[b]{0.45\linewidth}
\includegraphics[width = \linewidth]{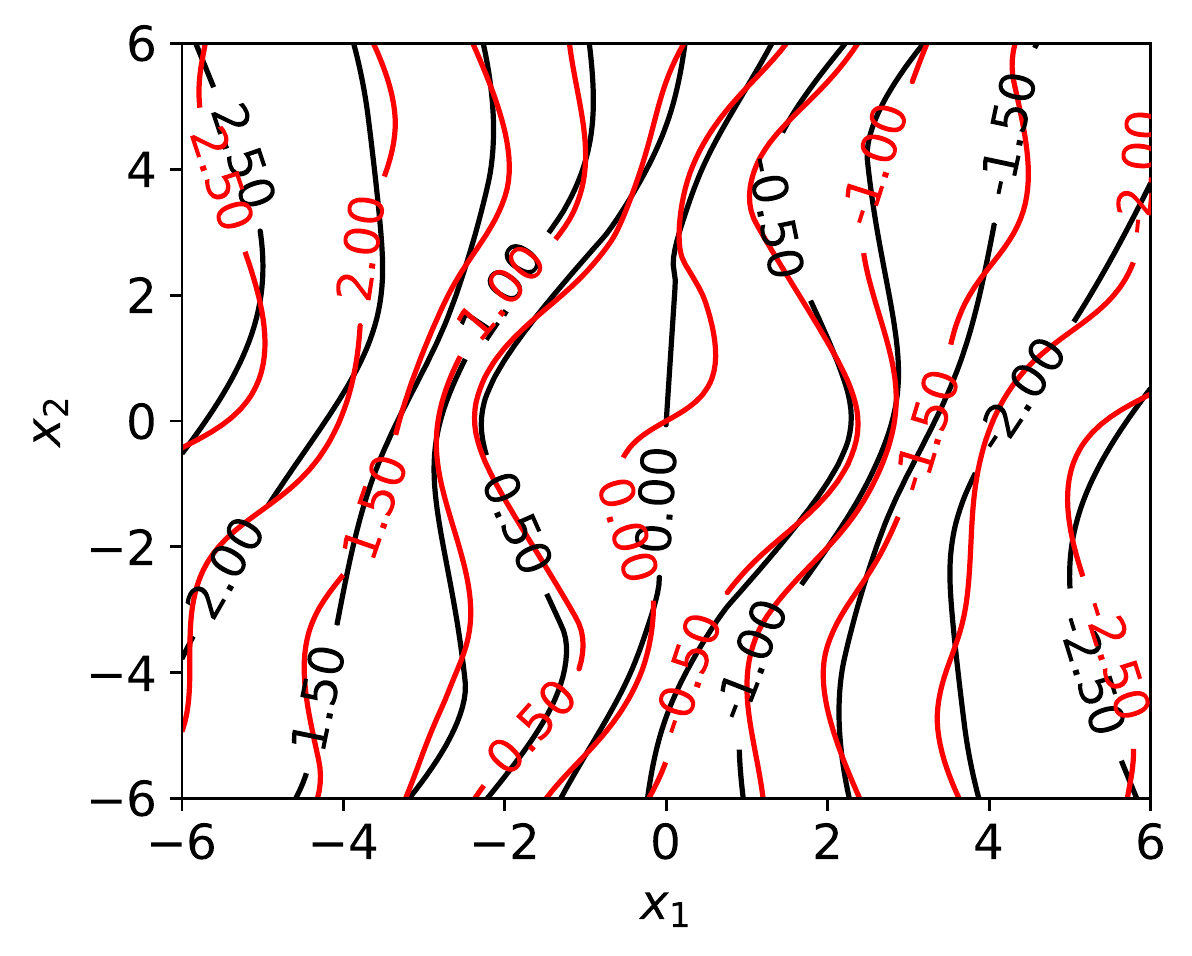}
\end{minipage}
\centering{\quad (c)}
\begin{minipage}[b]{0.47\linewidth}
\includegraphics[width = \linewidth]{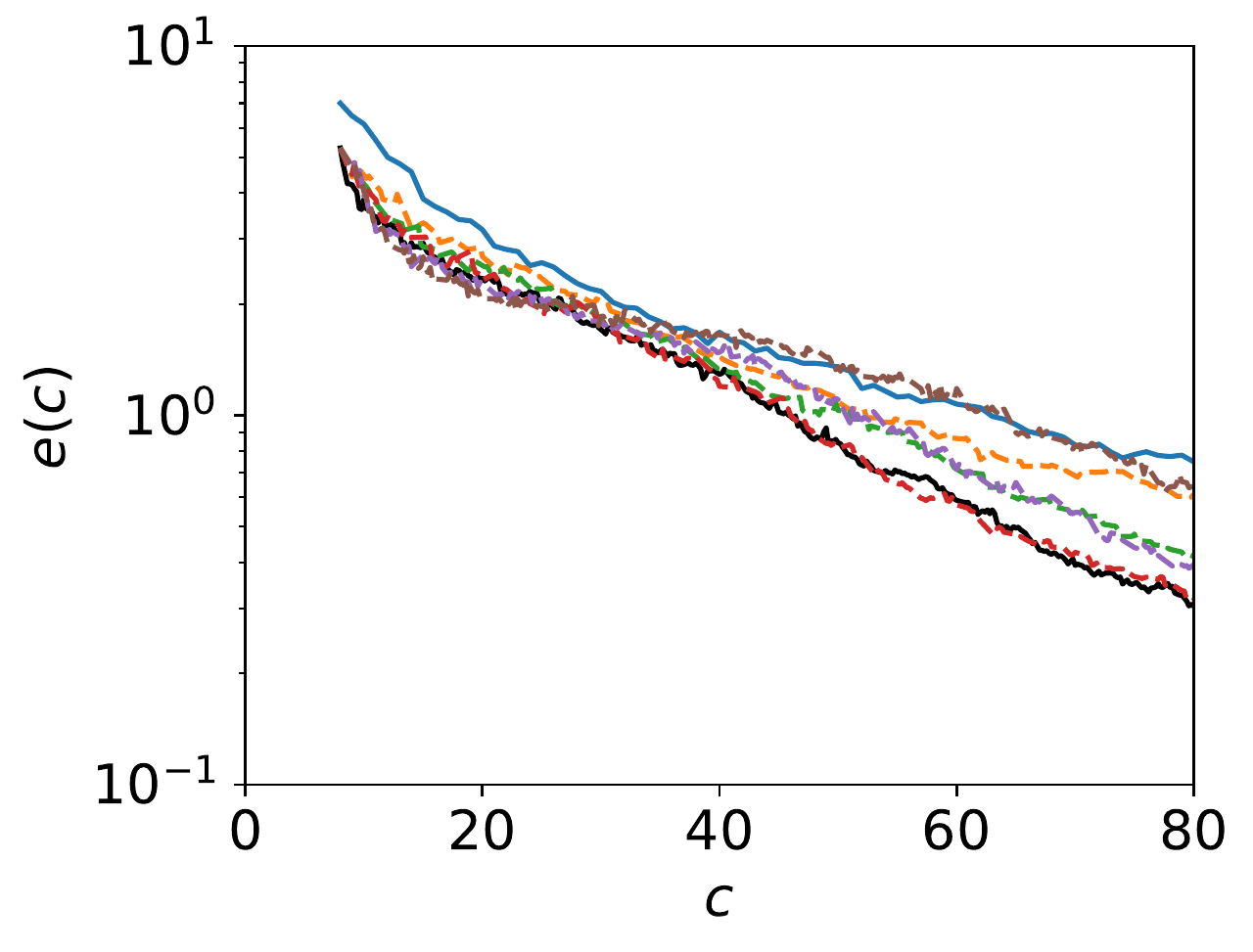}
\end{minipage}
\caption*{Figure \ref{fig:os_nonlinear_n}: See next page for caption.}
\end{figure}

\begin{figure}
\centering
\begin{minipage}[b]{0.45\linewidth}
\includegraphics[width = \linewidth]{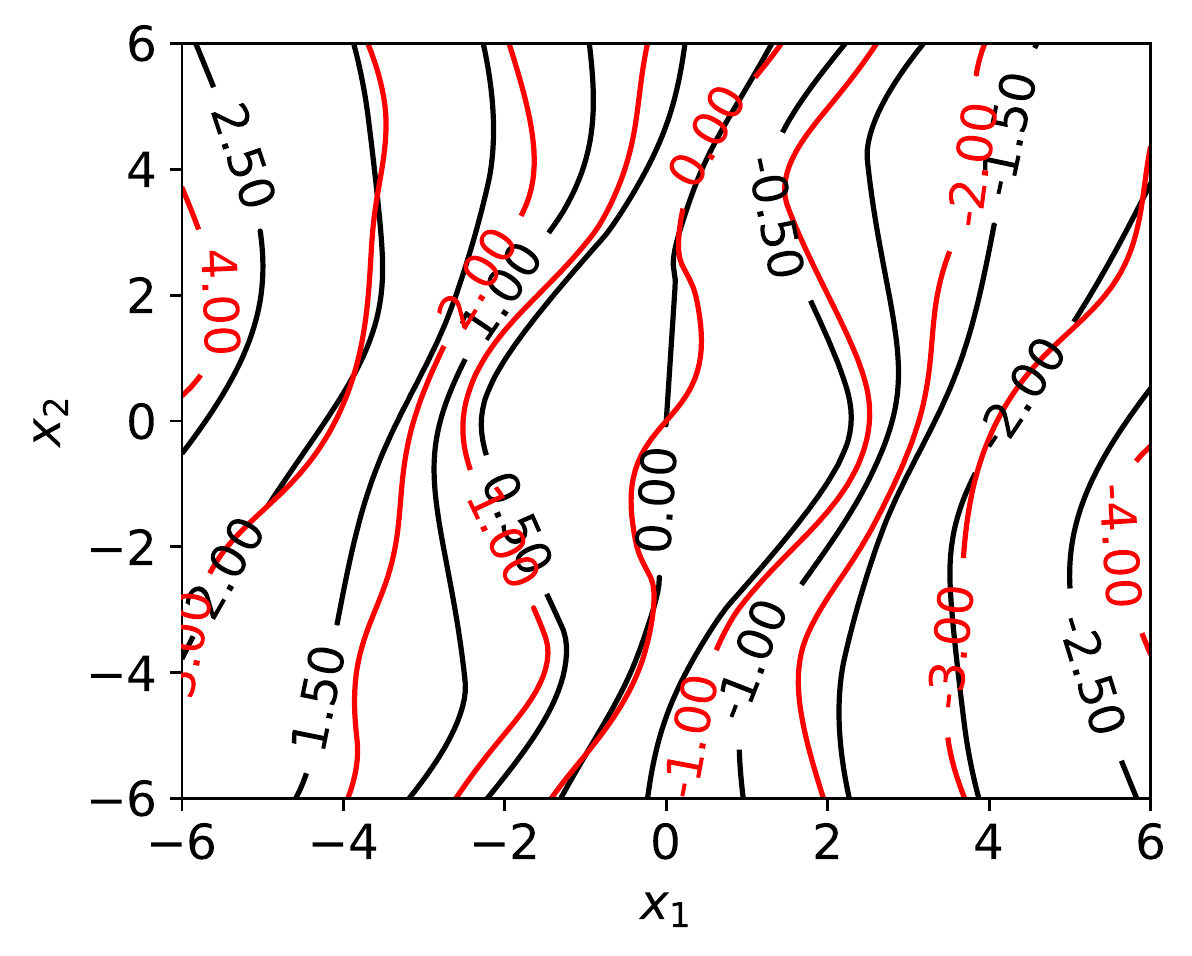}
\end{minipage}
\centering{\quad (d)}
\begin{minipage}[b]{0.47\linewidth}
\includegraphics[width = \linewidth]{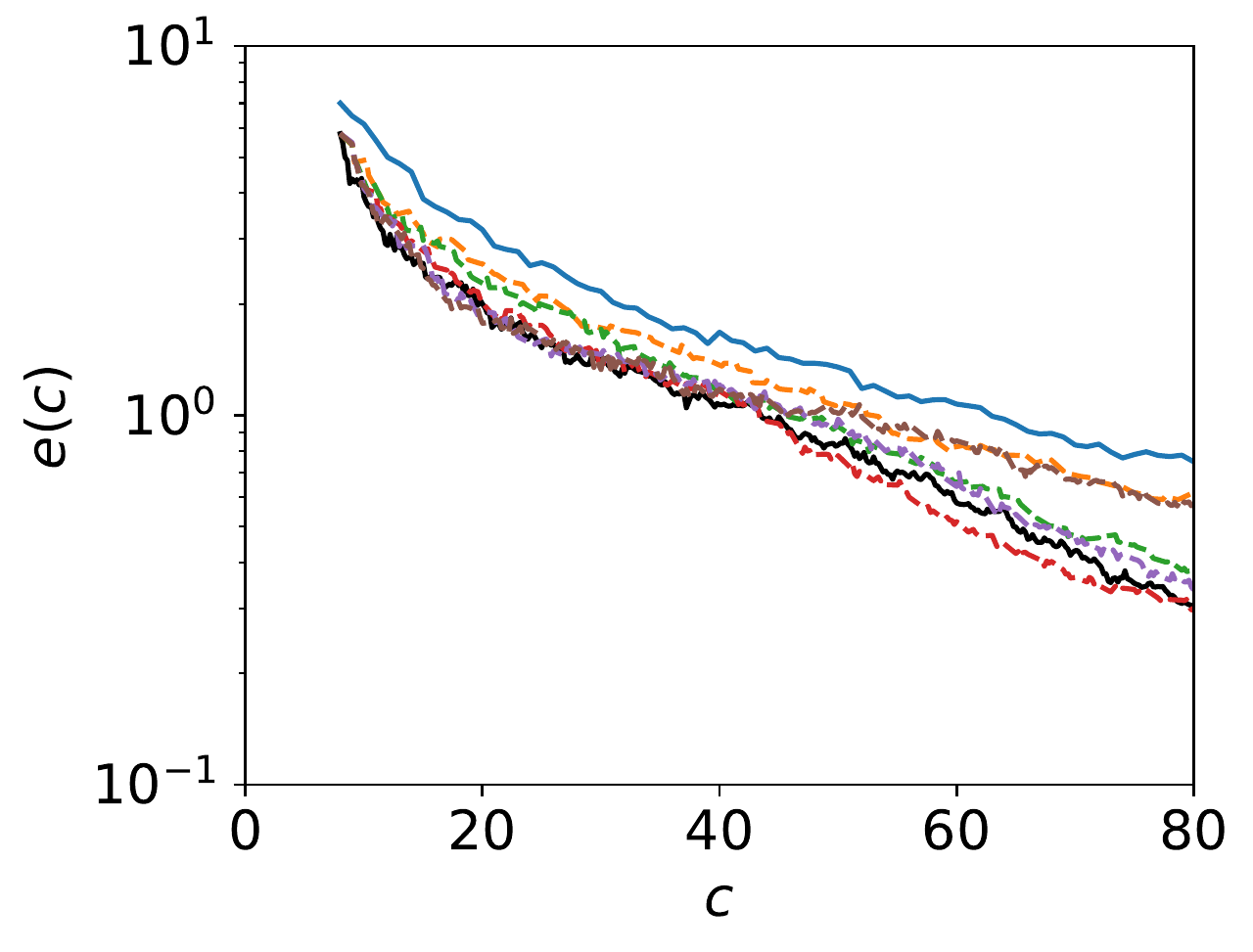}
\end{minipage}
\begin{minipage}[b]{0.45\linewidth}
\includegraphics[width = \linewidth]{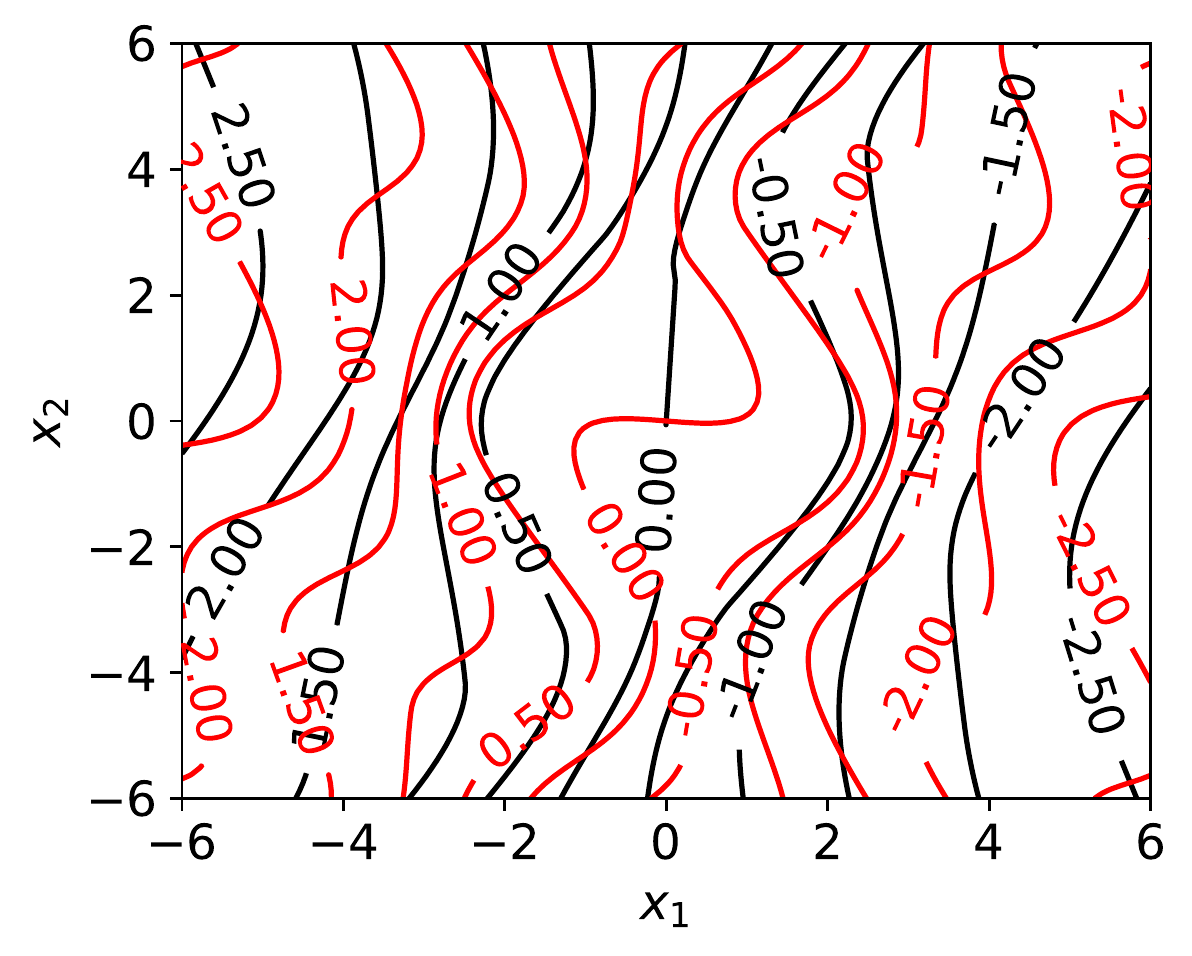}
\end{minipage}
\centering{\quad (e)}
\begin{minipage}[b]{0.47\linewidth}
\includegraphics[width = \linewidth]{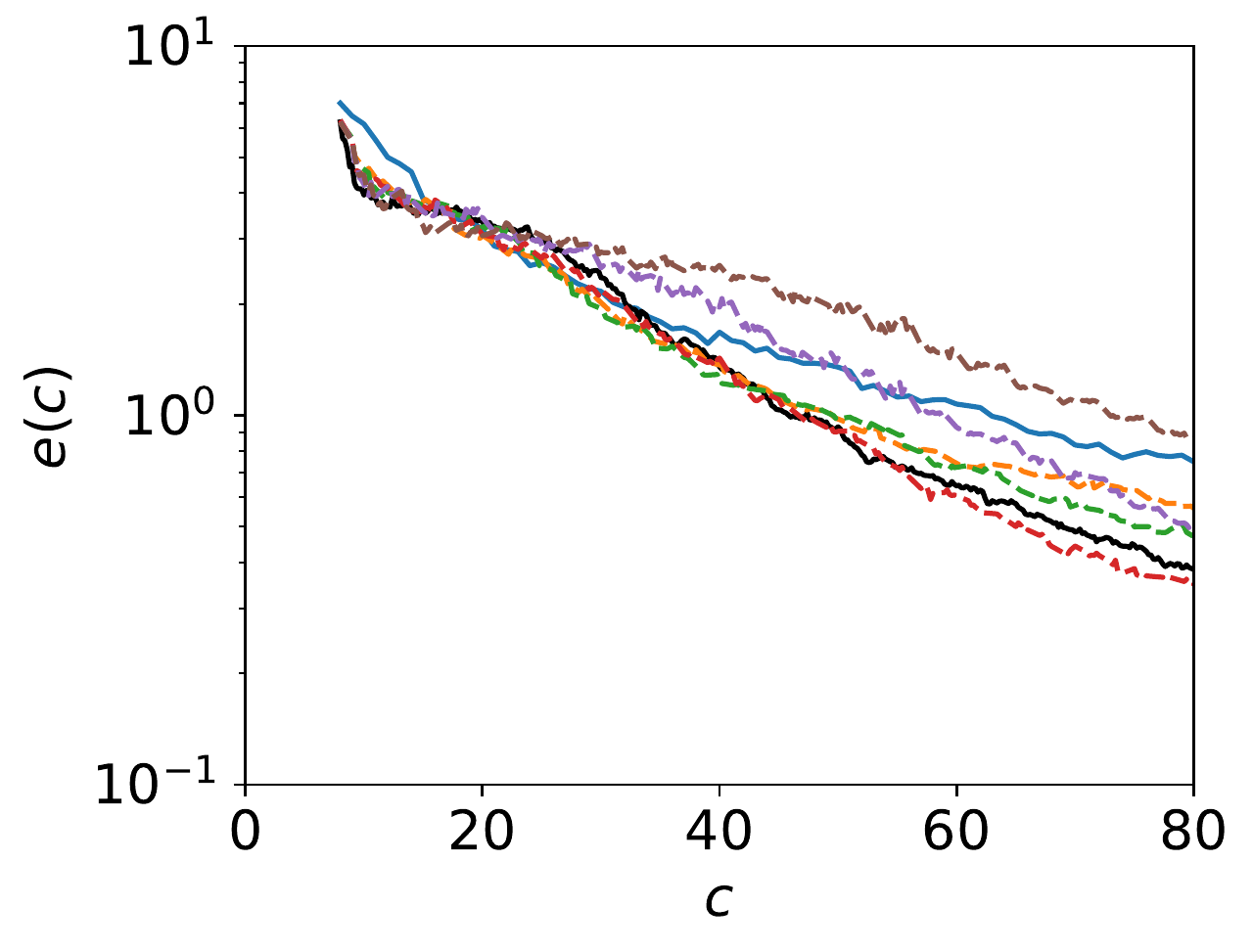}
\end{minipage}
\begin{minipage}[b]{0.45\linewidth}
\includegraphics[width = \linewidth]{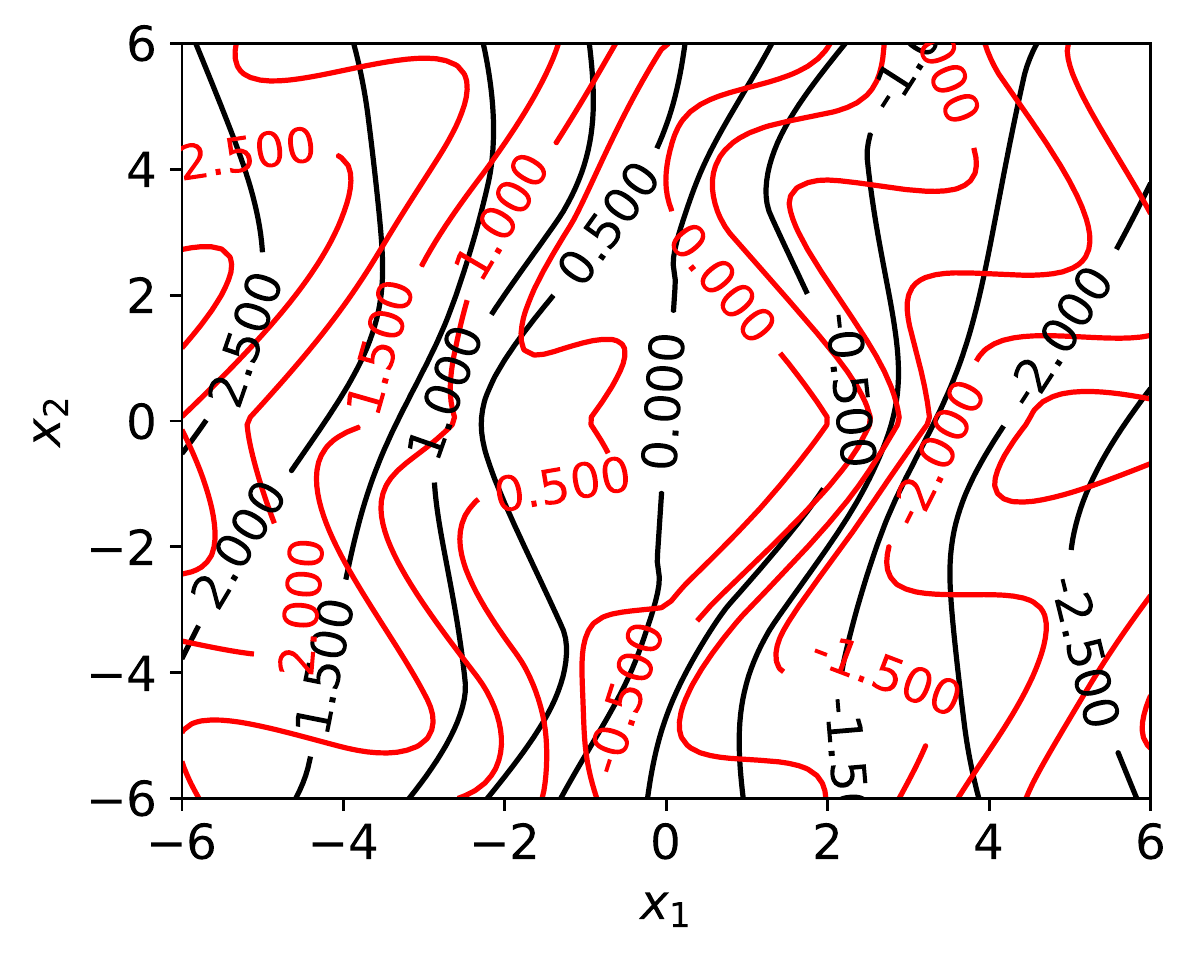}
\end{minipage}
\centering{\quad (f)}
\begin{minipage}[b]{0.47\linewidth}
\includegraphics[width = \linewidth]{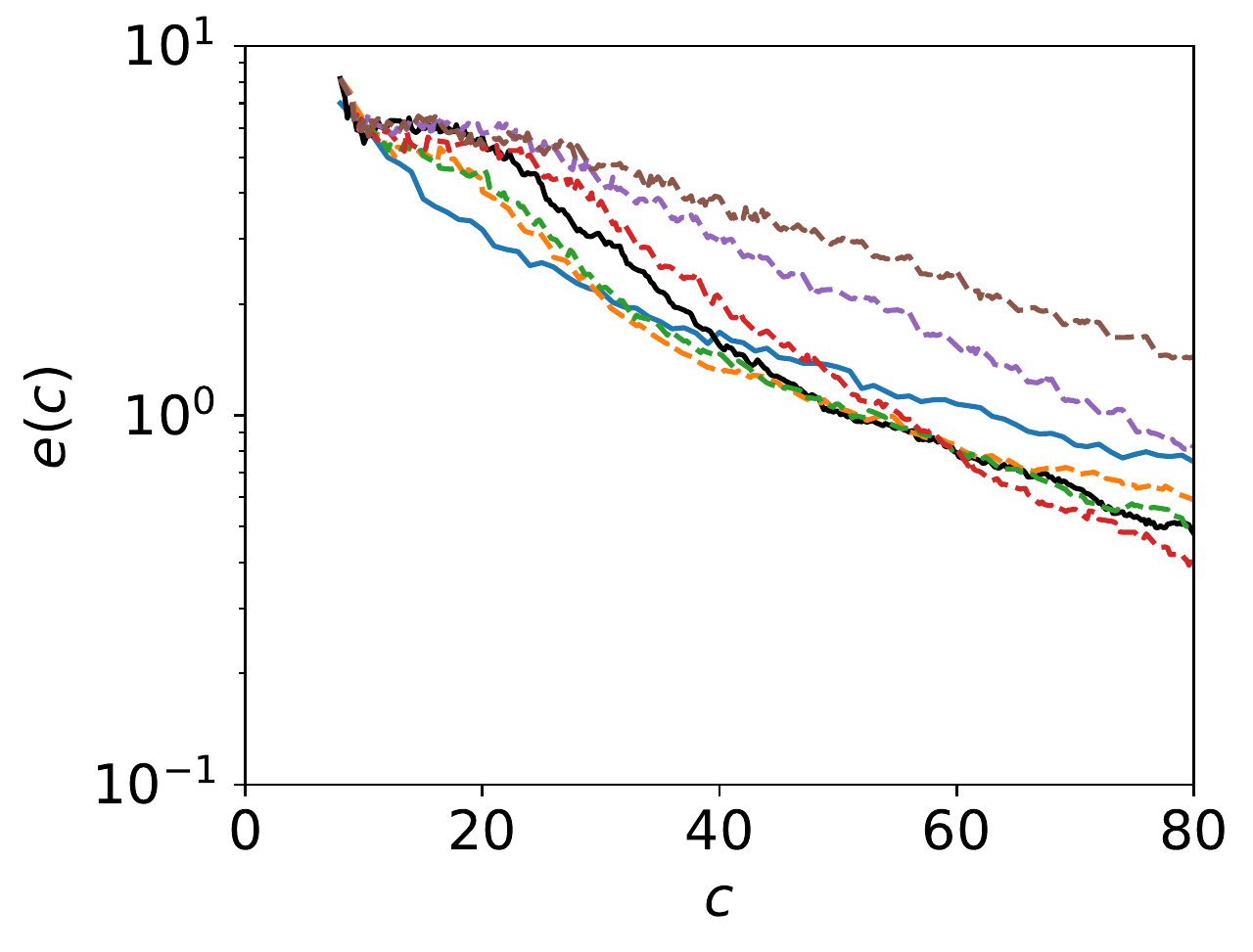}
\end{minipage}
\caption{Results of (a) case 2 $\{\delta=0.02, \rho=1\}$, (b) case 3 $\{\delta=0.05, \rho=1\}$, (c) case 5 $\{\delta=0.1, \rho=1\}$, (d) case 10 $\{\delta=0.1, \rho=1.5\}$, (e) case 8 $\{\delta=0.2, \rho=1\}$, and (f) case 9 $\{\delta=0.4, \rho=1\}$ for low-fidelity function \eqref{2d_low} with nonlinear difference in the 2D oscillator problem. Left: the low fidelity function (\redline) and high-fidelity function (\blackline). Right: the error $e(c)$ computed by BF-O(\blackline), BF-F1(\orangedashedline), BF-F2(\greendashedline), BF-F5(\reddashedline), BF-F10(\purpledashedline), BF-F15 (\browndashedline) and SF(\blueline).}
\label{fig:os_nonlinear_n}
\end{figure}

\begin{figure}
\centering
\begin{minipage}[b]{0.48\linewidth}
\includegraphics[width = \linewidth]{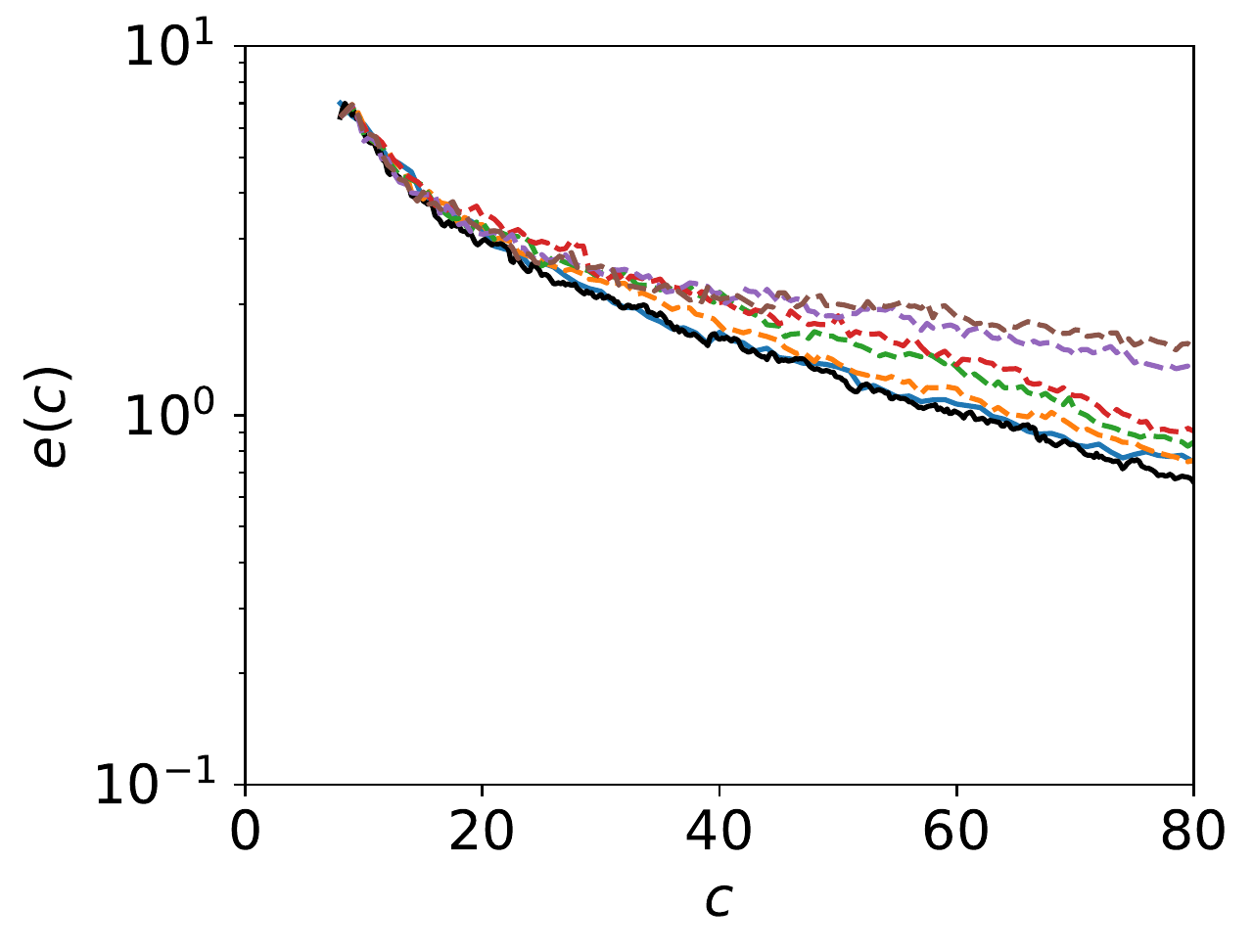}
\centering{\quad (a)}
\end{minipage}
\begin{minipage}[b]{0.48\linewidth}
\includegraphics[width = \linewidth]{figure/oscillator/errors/nonlinear_c5.pdf}
\centering{\quad (b)}
\end{minipage}
\begin{minipage}[b]{0.48\linewidth}
\includegraphics[width = \linewidth]{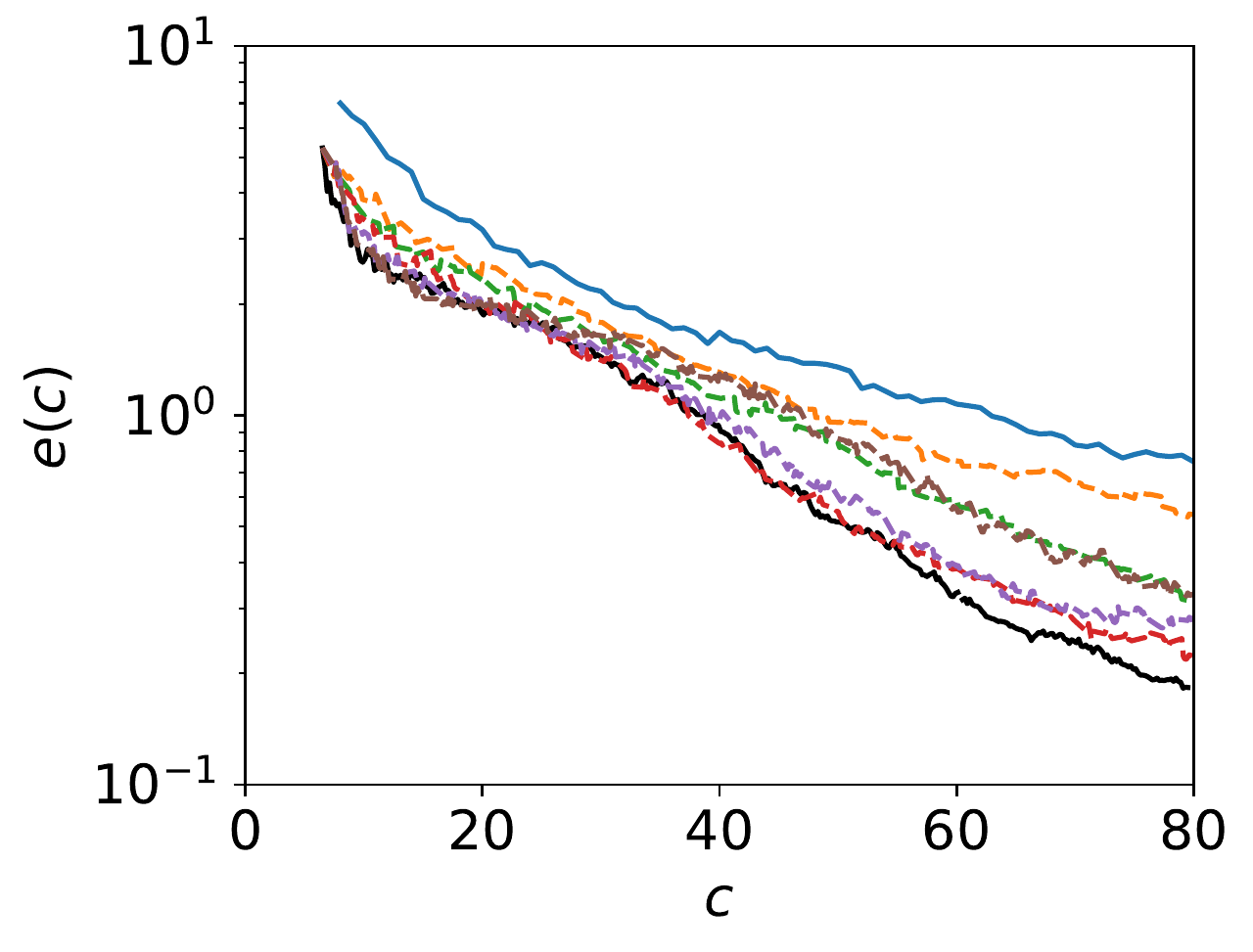}
\centering{\quad (c)}
\end{minipage}
\begin{minipage}[b]{0.48\linewidth}
\includegraphics[width = \linewidth]{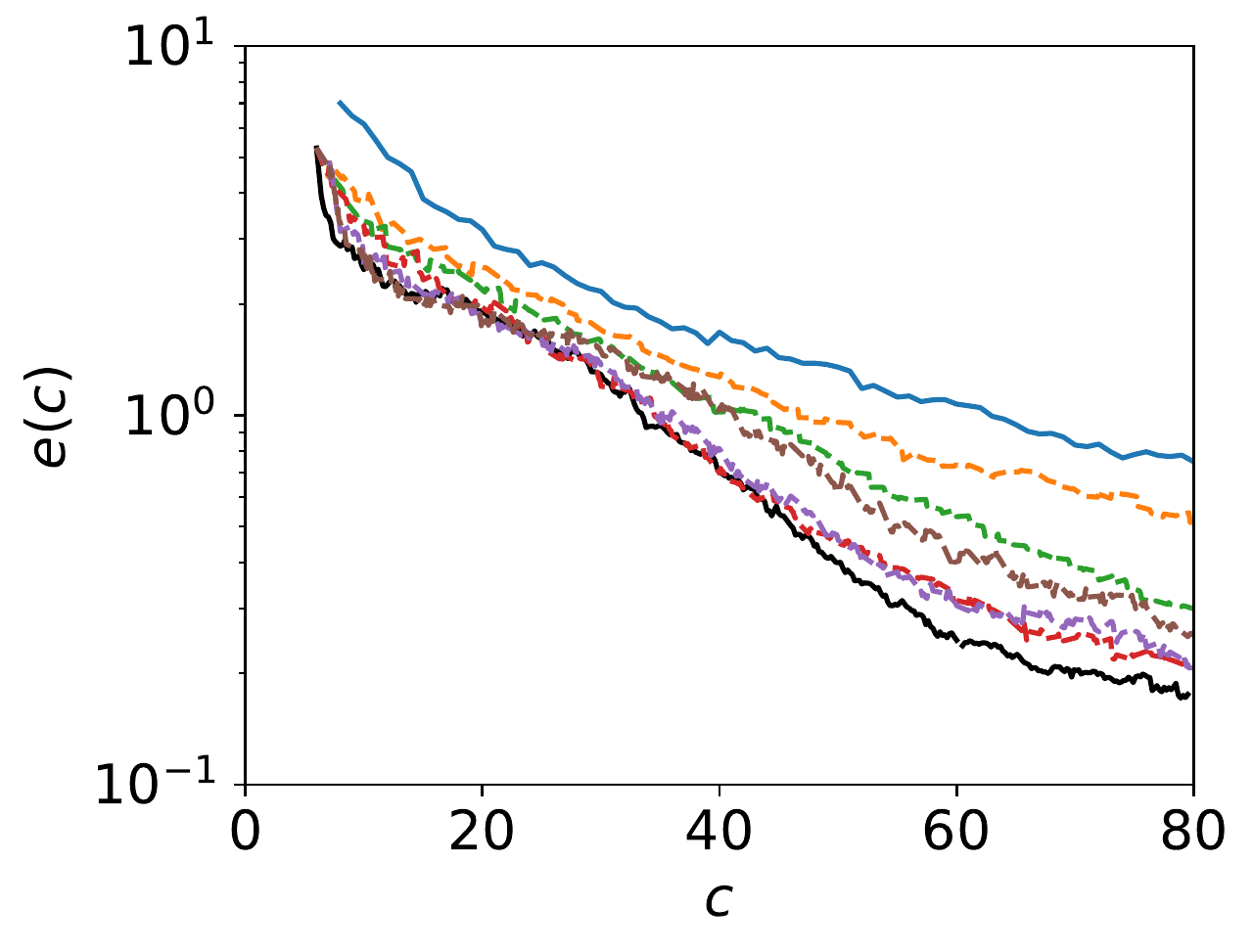}
\centering{\quad (d)}
\end{minipage}
\caption{Errors in the cases of $\{\delta=0.1, \rho=1\}$ with (a) case 4, $c_h/c_l=2$, (b) case 5, $c_h/c_l=5$, (c) case 6, $c_h/c_l=8$, and (d) case 7, $c_h/c_l=10$ computed by BF-O(\blackline), BF-F1(\orangedashedline), BF-F2(\greendashedline), BF-F5(\reddashedline), BF-F10(\purpledashedline), BF-F15 (\browndashedline) and SF(\blueline).}
\label{fig:os_nonlinear_c}
\end{figure}

To further test the performance of BF-O (as well as BF-F$n$) in more diversified situations, in the following, we construct additionally 9 cases with different computational costs $c_l$ and low-fidelity functions $f_l(\mathbf{x})$ as summarized in Table \ref{table:2d}. In addition to the linear difference function used in the previous case (now case 1 in Table \ref{table:2d}), we also consider a nonlinear difference function $d(\mathbf{x})=\delta \sin(x_1 + x_2)$ which is known to be a difficult situation to approximate by a Gaussian process \cite{rasmussen2003gaussian}.

Figure \ref{fig:os_nonlinear_n} shows the results for varying the low-fidelity accuracy level with fixed $c_h/c_l=5$, i.e., cases 2, 3, 5, 8-10 in Table \ref{table:2d} with varying $\rho$ and $\delta$, with the range of $\delta$ corresponding to difference terms up to 20\% of the maximum response $f_h(\mathbf{x})$ for $|\mathbf{x}|<4$. Together with figure \ref{fig:os_linear}, we see that the BF-O scheme consistently performs among the best of all tested methods. More specifically, with the increase of complexity in the difference function (i.e., increasing $\delta$ but not much for increasing $\rho$), the performance of BF-O can deteriorate for smaller number of sequential samples (e.g., figure \ref{fig:os_nonlinear_n}(f)), but still behaves close to the optimal at larger number of samples with cost $c\approx 80$. These cases correspond to the situation where the difference terms are difficult to learn, which takes more cost to make low-fidelity samples useful for the final results. 

Figure \ref{fig:os_nonlinear_c} shows the results for varying $c_l$, i.e., $c_h/c_l=2,5,8,10$ with fixed $c_h=1$ as in cases 4-7 in Table \ref{table:2d}. Similar to the results above, we see that the BF-O method consistently performs among the best. This indicates that the acquisition function \eqref{tfgp_acq} employing the ratio between benefit $B(i,\mathbf{x})$ (uncertainty reduction) and cost $c_i$ effectively captures the optimal that balances the two factors. 

The performance of the BF-O scheme, in terms of the its ranking in all schemes, is summarized in Table \ref{table:2d}. It is clear that the BF-O method consistently provides accurate results of the extreme-value response PDF (ranking the 1st or 2nd among all methods for all cases). Moreover, we include in the table the average ratio of low and high fidelity sample numbers $n_l/n_h$ in the BF-O method, as well as the value of $n$ corresponding to the BF-F$n$ method with the best performance at $c=80$. While the optimal $n$ in the BF-F$n$ method does not necessarily correspond to the $n_l/n_h$ in BF-O method, we find that the two numbers are close in most cases, with the latter automatically captured by the algorithm.


\subsection{High-dimensional borehole hydrological model}
\begin{table}
\caption{The input parameters and their distributions of the borehole function.}
\begin{threeparttable}[t]
\begin{tabular}{|c|c|c|c|} 
\hline
    Parameter & Definition & Range & Distribution.\\ 
\hline
    $r_w$ & radius of borehole & [0:05; 0:15] $m$ & normal \tnote{1} \\ 
    $r$ & radius of influence & [100; 50000] $m$ & logNormal \tnote{2}  \\ 
    $T_u$ & transmissivity of upper aquifer & [63070; 115600] $m^2/yr$ & uniform  \\ 
    $H_u$ & potentiometric head of upper aquifer & [990; 1110] $m$ & uniform  \\ 
    $T_l$ & transmissivity of lower aquifer & [63:1; 116] $m^2/yr$ & uniform  \\ 
    $H_l$ & potentiometric head of lower aquifer & [700; 820] $m$ & uniform  \\ 
    $L$ & length of borehole & [1120; 1680] $m$ & uniform \\ 
    $k_w$ & hydraulic conductivity of borehole & [9855; 12045] $m/yr$ & uniform  \\ 
\hline   
\end{tabular}
\begin{tablenotes}
     \item[1] $r_w$ with mean 0.10, stand deviation 0.0161812.
     \item[2] $\ln(r)$ with mean 7.71, stand deviation 1.0056.
\end{tablenotes}
\end{threeparttable}
\label{table:2}
\end{table}
We next consider an eight-dimensional borehole hydrological model, which has been used as an example of high-dimensional problems to quantify the extreme response statistics using single-fidelity methods \cite{harper1983sensitivity,blanchard2021bayesian, farazmand2017variational}. The model physically computes the flow rate through a borehole, formulated as
\begin{equation}
    f_h(\mathbf{x}) = \frac{2 \pi T_u(H_u-H_l)}{\ln (r/r_w)}[1 + \frac{2LT_u}{\ln(r/r_w)r^2_w K_w}
    + \frac{T_u}{T_l} ],
\label{borehole_h}
\end{equation}
with an eight-dimensional input $\mathbf{x} = \{r_w, r, T_u, H_u, T_l, H_l, L, K_w\}$, including their distributions, detailed in Table \ref{table:2}. We further construct a low-fidelity model as 
\begin{equation}
    f_l(\mathbf{x}) = f_h(\mathbf{x}) + 7.5 \sin(x_1) + 0.75 \sin(\sum_{i=2}^8 x_i),
\label{borehole_l}
\end{equation}
where we use a nonlinear (sinusoidal) form of the difference function and put more weights on the parameter $x_1=r_w$, as it is the most influential factor to the response $f_h(\mathbf{x})$ \cite{harper1983sensitivity}. In particular, the coefficient 7.5 is chosen such that the term of $\sin(x_1)$ corresponds to 5\% of the maximum response. We keep the computational cost ratio as $c_h/c_l=5$ for this case.

\begin{figure}
    \centering
    \includegraphics[width=0.48\linewidth]{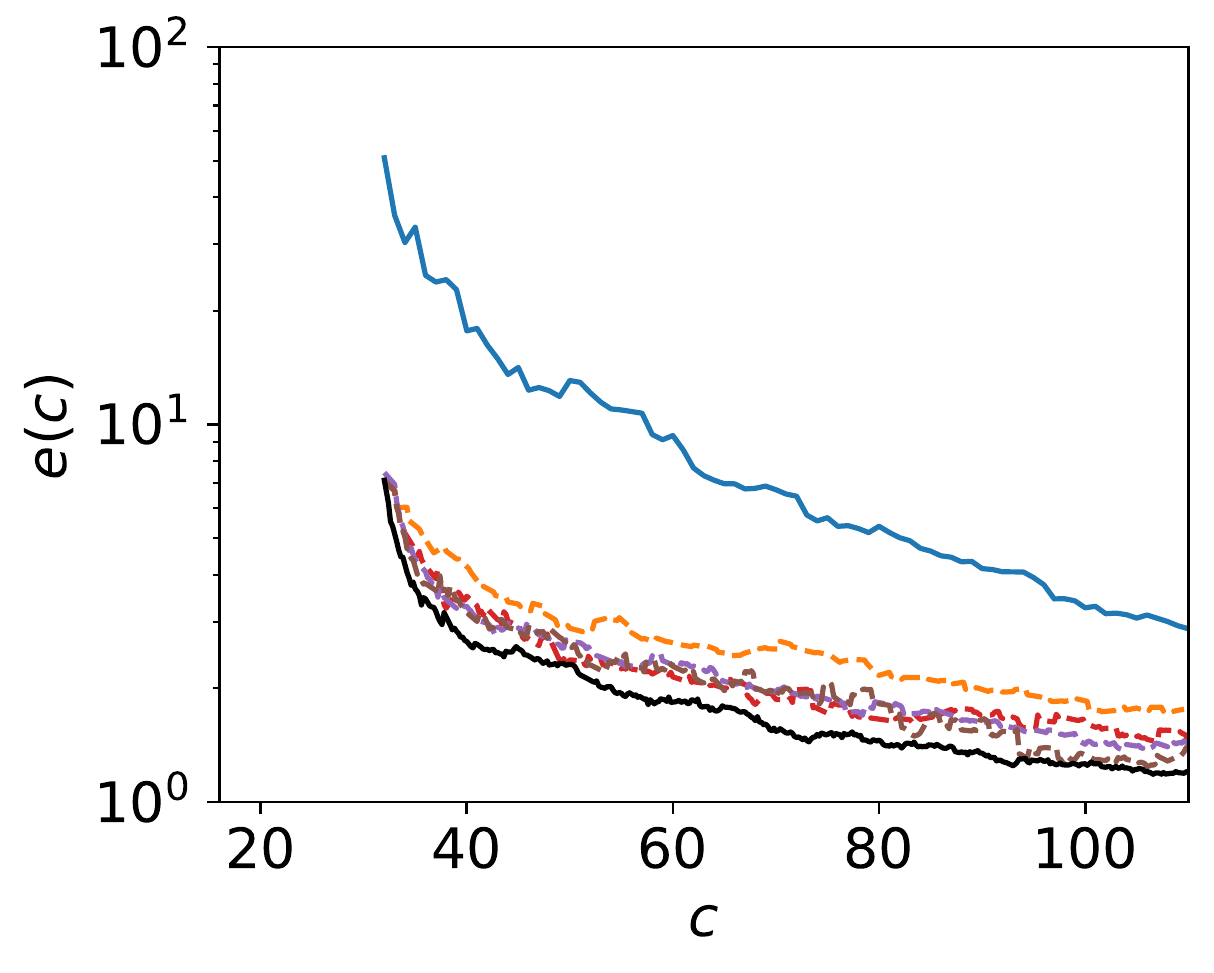}
    \caption{The error $e(c)$ computed by BF-O(\blackline), BF-F1(\orangedashedline), BF-F5(\reddashedline), BF-10(\purpledashedline), BF-F15 (\browndashedline) and SF(\blueline) for the high-dimensional borehole problem.}
    \label{fig:borehole}
\end{figure}

The results from the SF, BF-O and BF-F$n$ methods are shown in figure \ref{fig:borehole}.  Comparing to the SF result, it is clear that the benefit of using bi-fidelity models is more evident than the low-dimensional cases, even for the initial data set without sequential samples. The reason may lie in the fact that the ``value'' of a high-fidelity sample becomes compromised with the increase of dimensions. In addition, the BF-O method again performs the best among all BF-F$n$ methods with varying $n$.  

\begin{figure}
    \centering
    \includegraphics[width=8cm]{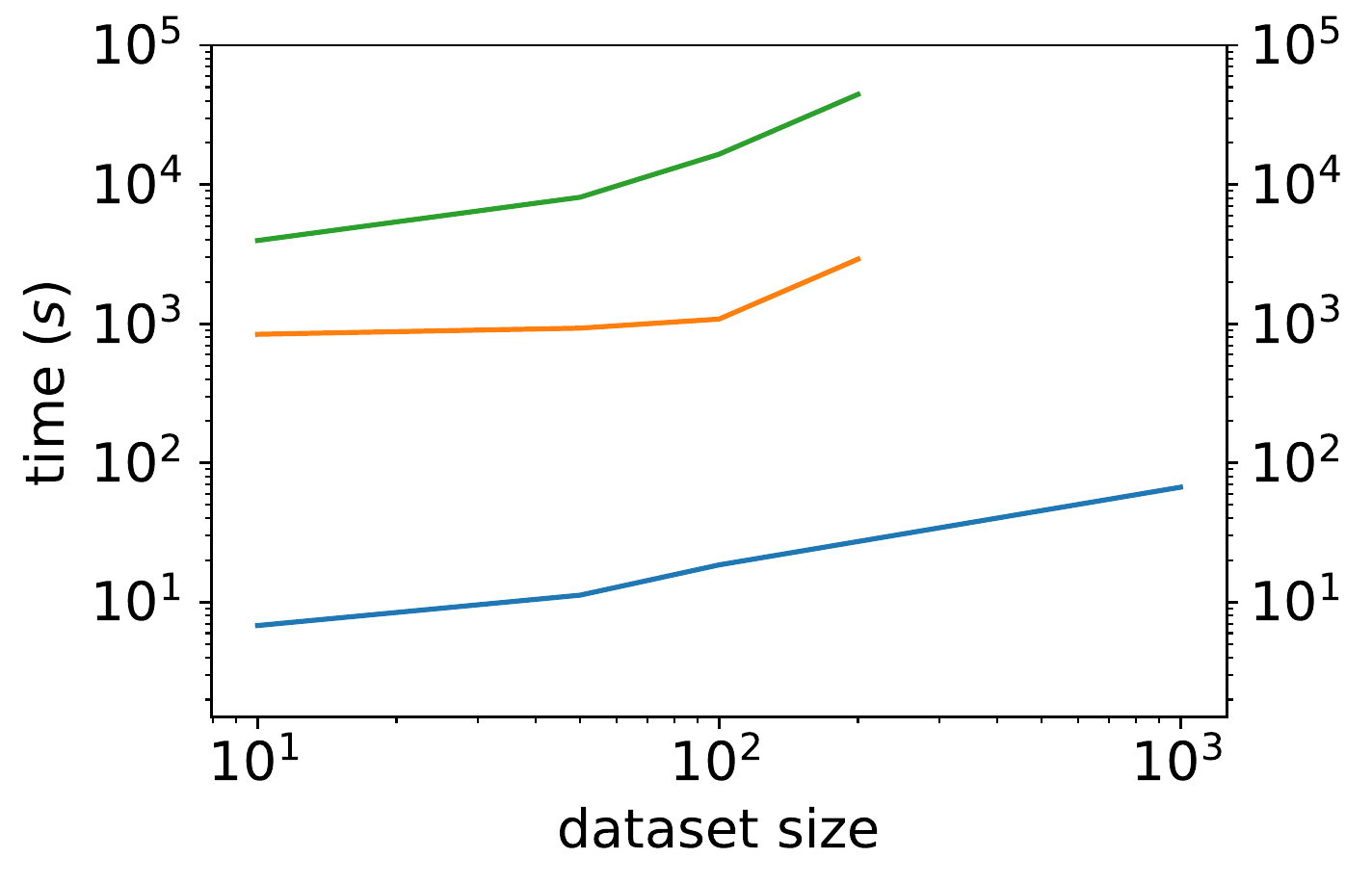}
    \caption{The computation time for selecting one sequential sample in the borehole problem by using the $U$ criterion in \eqref{acq_L}(\greenline), the $B$ criterion in \eqref{tfgp_acq} with numerical integration (\orangeline) and analytical formula (\blueline) as objective functions for different sizes of the existing dataset. In the former two computations, we assume that 10 times (a conservative number) of acquisition evaluations are needed compared to the gradient-based optimization in the third case, as a common practice found in \cite{martins2021engineering}.}
    \label{fig:time}
\end{figure}

We finally remark that the computation in this eight-dimensional case is only enabled because of (i) the development of \eqref{tfgp_acq} which avoids the construction of a new Gaussian process for each hypothetical sample, and (ii) the development of analytical formula through GMM which avoids the numerical integration in \eqref{tfgp_acq} and enables the gradient computation. To illustrate this point, we show in figure \ref{fig:time} the computation time for solving \eqref{opt} as a function of the number of samples in the existing dataset on a single core of Intel Xeon Gold 6154 CPU (specifically we use $n_{GMM}=2$ with \eqref{gmm} computed by $10^6$ quadrature points, and 10 starting points in the quasi-Newton method). For comparison, we also include in figure \ref{fig:time} the computation time of using \eqref{acq_L} (with a new Gaussian process for each hypothetical point) and \eqref{tfgp_acq} (with numerical integration) as objective functions in optimization. In both cases, not only are the computation for the objective functions expensive, these computations also need to be repeated many more times in gradient-free optimization than that in the gradient-based method, resulting in prohibitive computational costs for large number of samples. In contrast, using the analytical formula combined with the GMM model, the computation takes only O(100)s even for a dataset of 1000 sample points, which is supposed to be negligible compared to the evaluation of the output of the high-fidelity physical model. 


\subsection{Coupling to CFD to compute extreme ship motion statistics in waves}
\begin{figure}
    \centering
    \includegraphics[width=16cm]{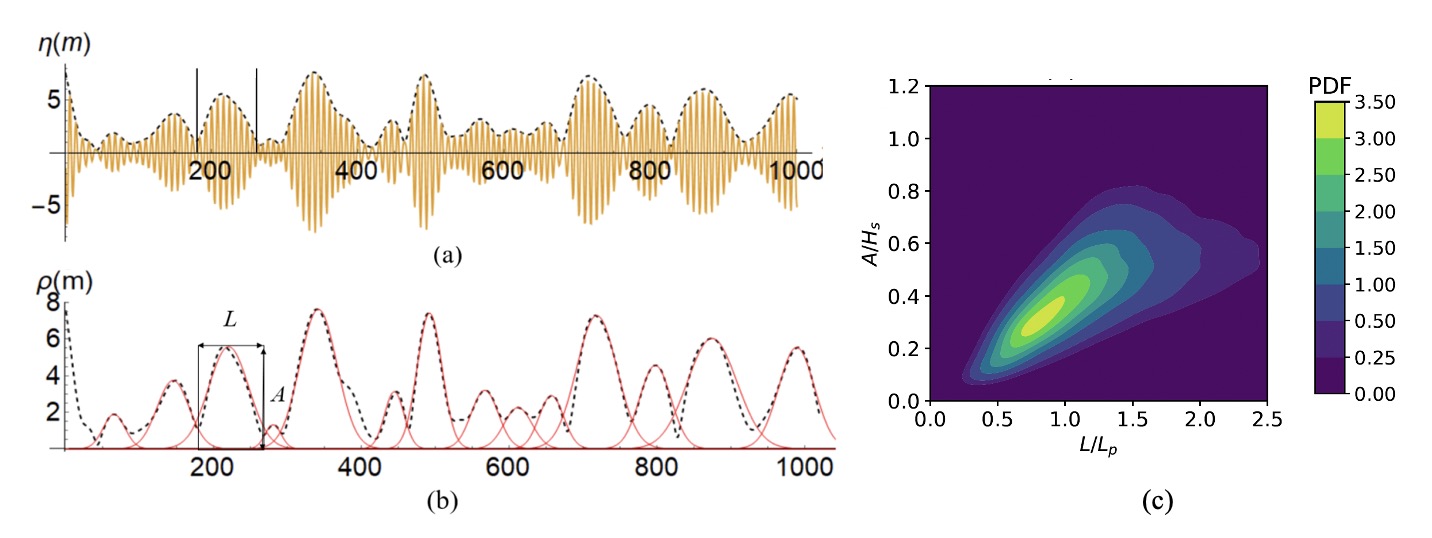}
    \caption{An example of a wave field with (a) wave elevation $\eta$ (\orangeline) and (b) envelop $\rho$ (\blackdashedline) approximated by a sequence of wave groups (\redline) with group amplitude parameter $A$ and length parameter $L$. (c) Joint PDF of $A$ (normalized by the significant wave height $H_s$) and $L$ (normalized by the spectral peak wavelength $L_p$).}
    \label{fig:wave_group}
\end{figure}

We further consider an application of our method to evaluate the PDF of extreme ship roll motion in irregular waves. More specifically, we study the motion of a two-dimensional, square-shaped hull geometry with $40m\times40m$ cross section and density $\rho_h=0.5\rho_w$ (with $\rho_w$ the water density) subject to beam waves. The input to this problem is considered as $\mathbf{x} = \{A,L\}$, with $A$ and $L$ the wave group amplitude and length, as a reduced-order description of an uni-directional irregular wave field \cite{cousins2016reduced, gong2021full}. Figure \ref{fig:wave_group} shows an example to evaluate parameters $A$ and $L$ from a given wave field as well as the resulted probability distribution $p_\mathbf{x}(A,L)$. The irregular wave field is described by a Gaussian spectrum in the form
\begin{equation}
F(k)\sim \exp \frac{-(k-k_0)^2}{2\mathcal{K}^2},  
\label{spectrum}
\end{equation}
with the significant wave height $H_s=12m$, peak (carrier) wavenumber $k_0=0.018 m^{-1}$ (corresponding to peak period $T_p=15s$), and $\mathcal{K}=0.05k_0$. The response in this case is considered as the maximum ship roll $r_{max}$ in a wave group, i.e., we consider a response function $r_{max}(A,L)$.

\begin{figure}
    \centering
    \begin{minipage}[b]{0.48\linewidth}
    \includegraphics[width = 0.9\linewidth]{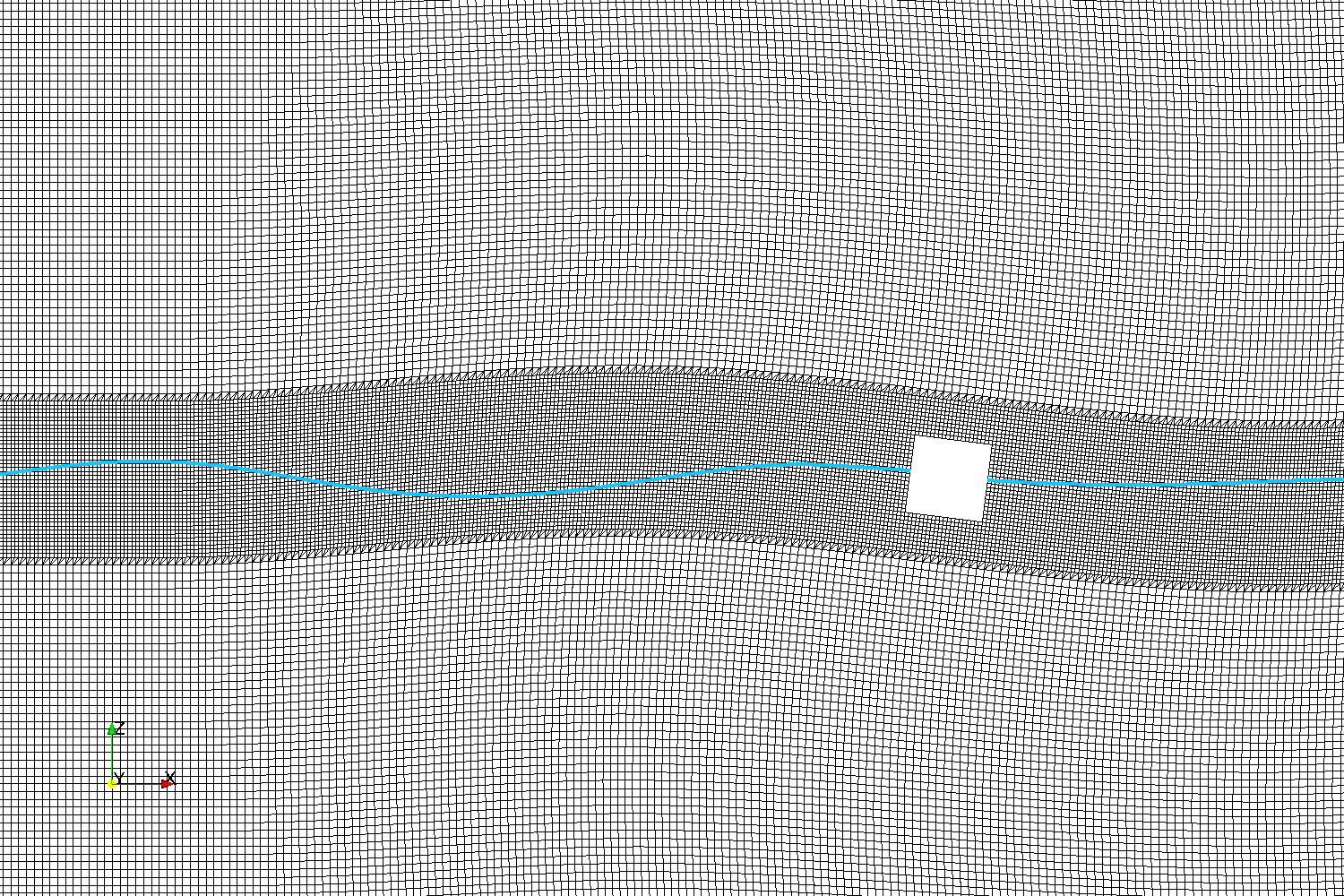}
    \centering{\quad (a)}
    \end{minipage}
    \begin{minipage}[b]{0.48\linewidth}
    \includegraphics[width = 0.9\linewidth]{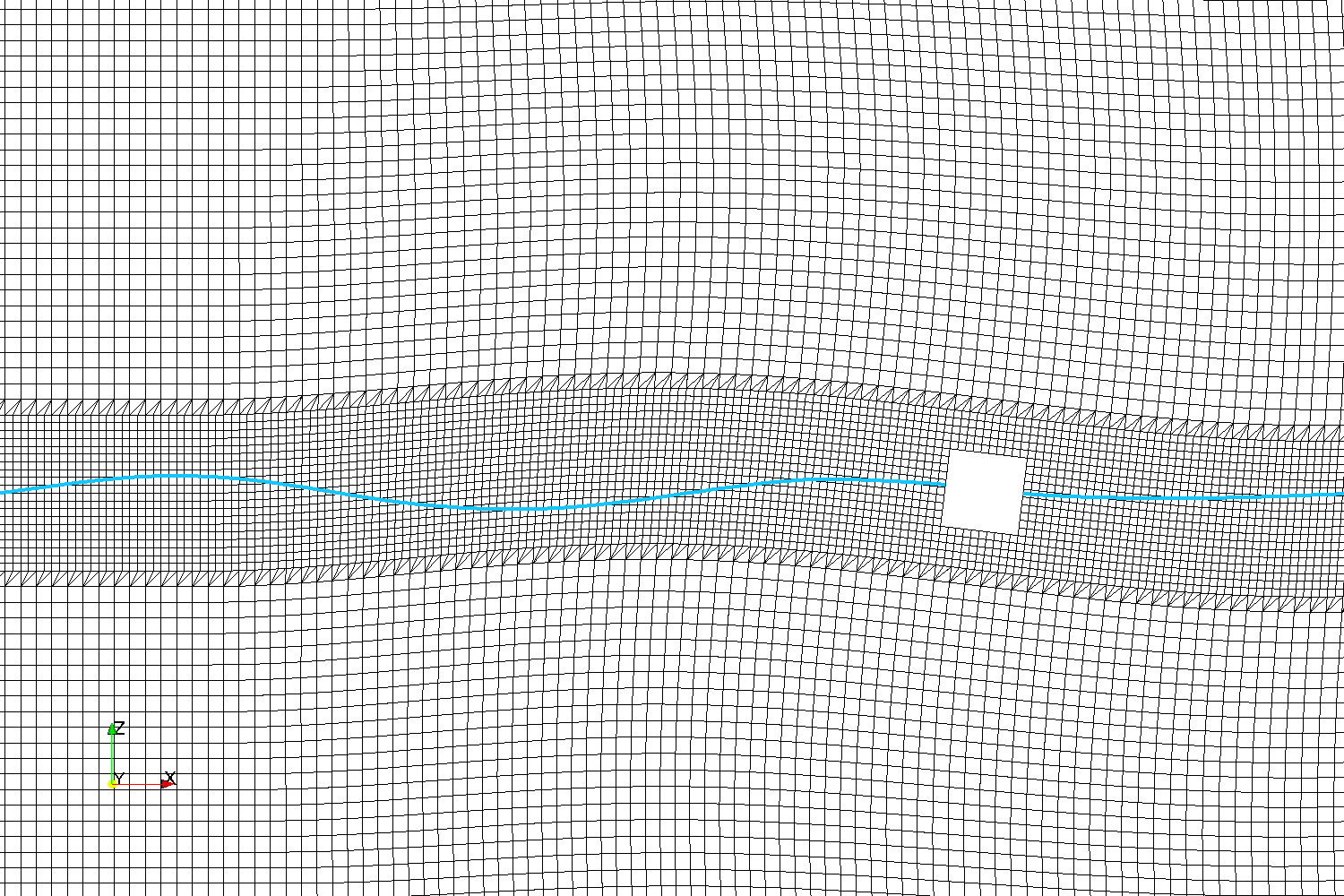}
    \centering{\quad (b)}
    \end{minipage}
    \centering
    \caption{(a) Fine and (b) coarse grids of the CFD simulations, with free surface indicated by a cyan line, and hull by a white box in each case.}
    \label{fig:cfd_schematic}
\end{figure}

\begin{figure}
    \centering
    \includegraphics[width=15cm]{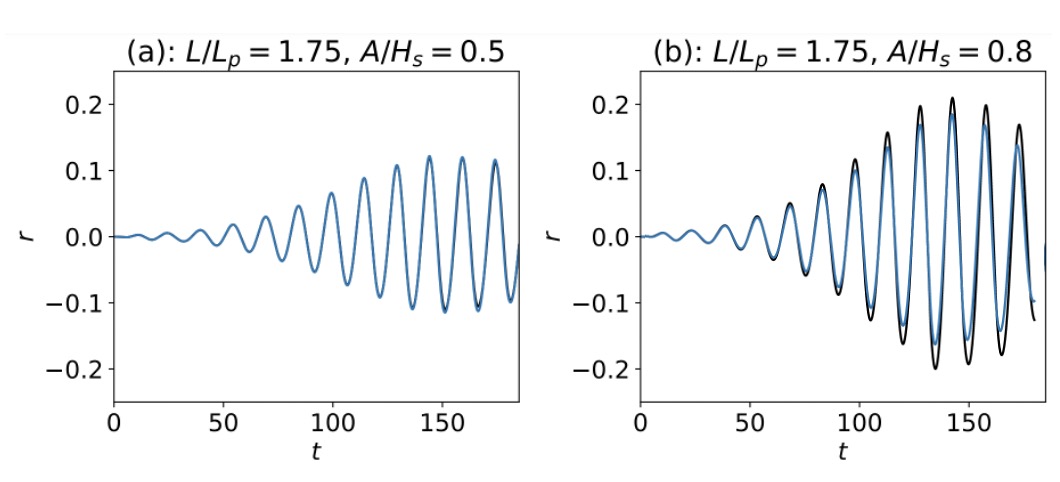}
    \caption{
    Results of high-fidelity (\blackline) and low-fidelity (\blueline) simulations for wave groups of (a) $L/L_p=1.75$, $A/H_s=0.5$ and (b) $L/L_p=1.75$, $A/H_s=0.8$.}
    \label{fig:cfd_results}
\end{figure}

\begin{figure}
\centering
\quad
\begin{minipage}[b]{0.443\linewidth}
\includegraphics[width = \linewidth]{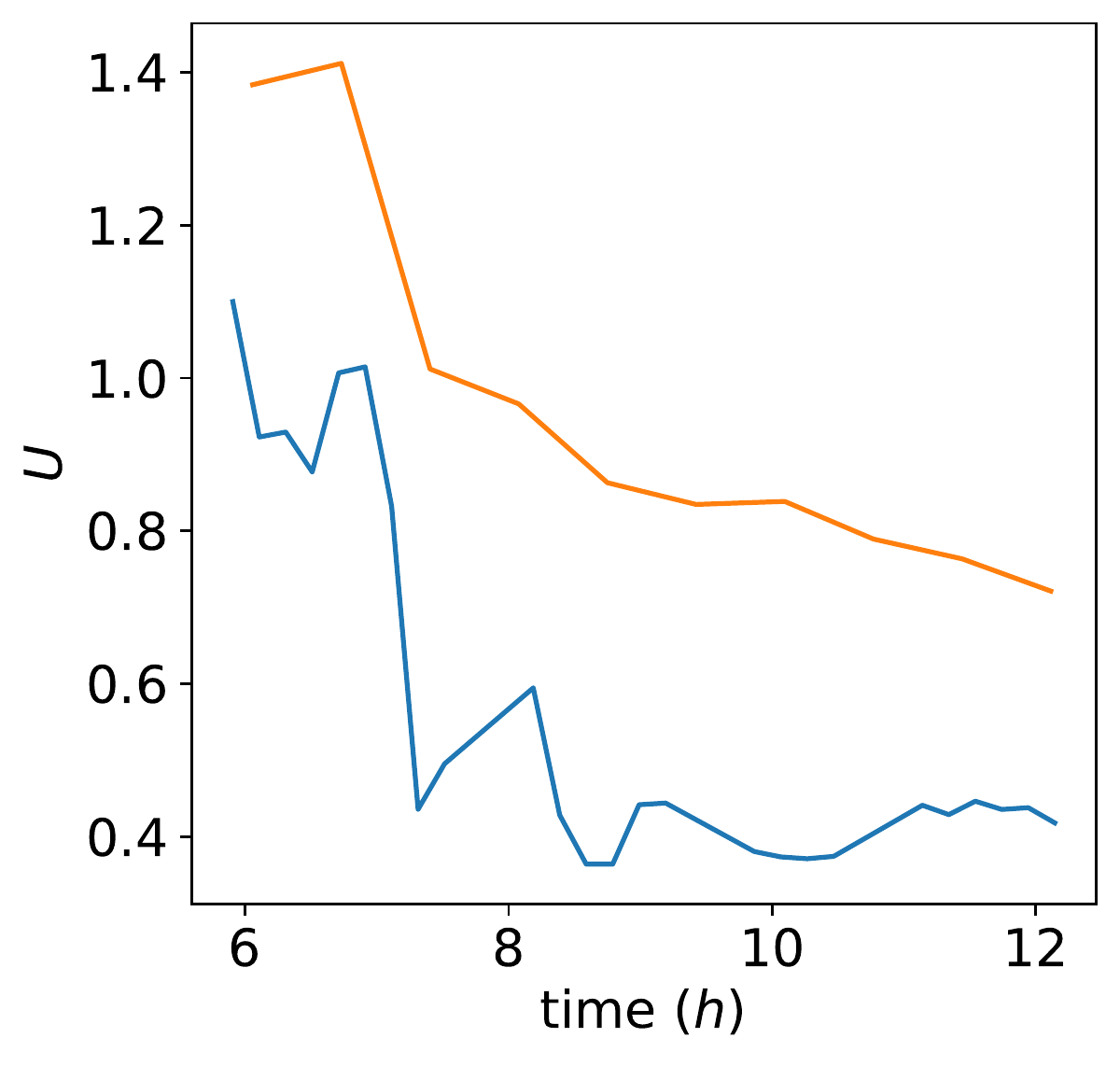}
\centering{\quad (a)}
\end{minipage}
\begin{minipage}[b]{0.46\linewidth}
\includegraphics[width = \linewidth]{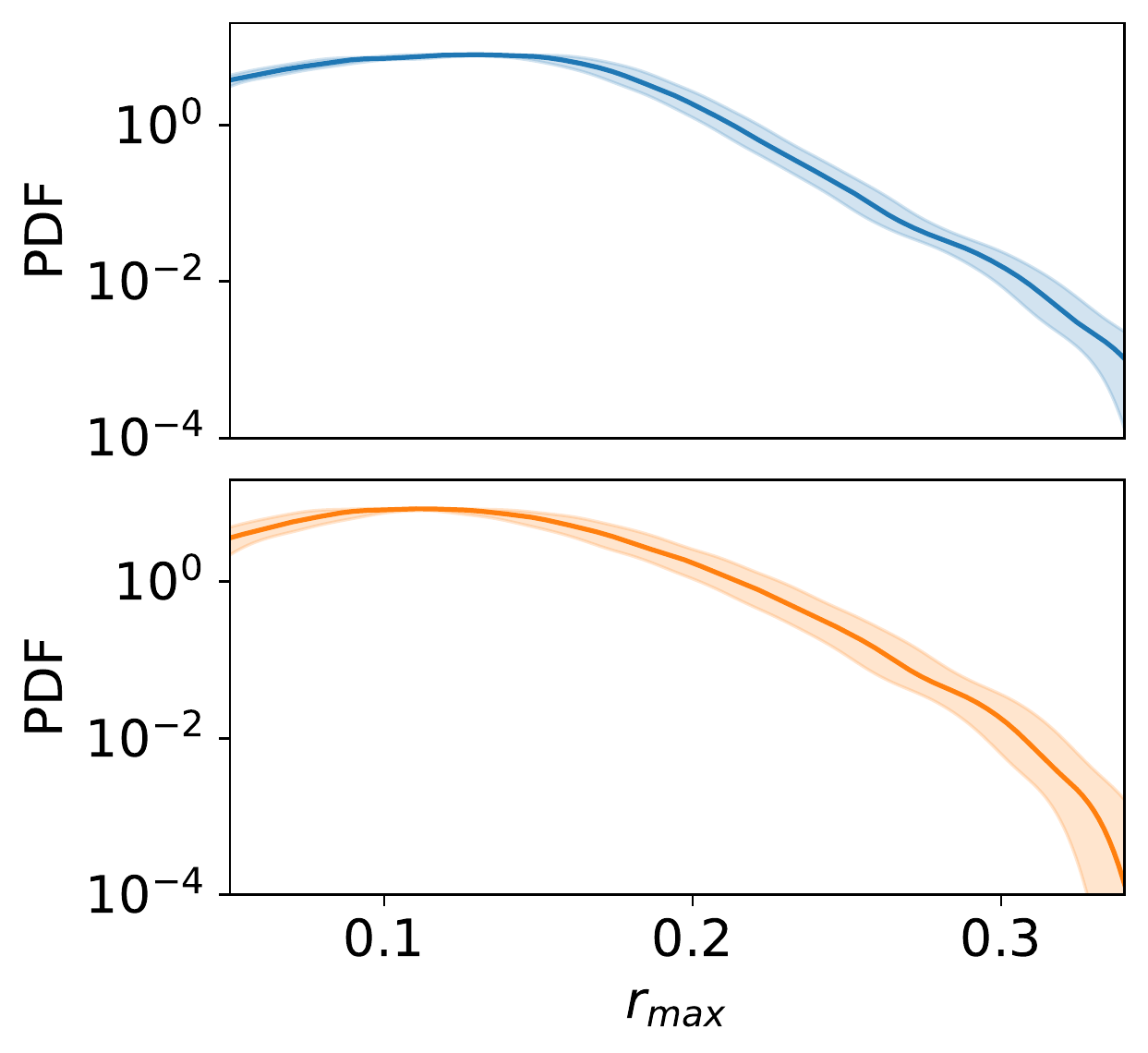}
\centering{\quad (b)}
\end{minipage}
\caption{(a) The uncertainty level $U$ computed by \eqref{acq_L} for BF-O (\blueline) and SF (\orangeline) as a function of the total computation time. (b) The final PDF computed by BF-O (\blueline, upper panel) and SF (\orangeline, lower panel) at approximately 12 hours of computation time, with two standard deviations marked by the shaded regions.}
\label{fig:ship_results}
\end{figure}

The high and low fidelity models in this case are constructed by CFD models with high and low resolutions, both developed using the open-source code OpenFOAM \cite{jasak2009openfoam}. In particular, the grid resolutions for both cases are shown in figure \ref{fig:cfd_schematic}, where the low-fidelity model uses half number of grids in both horizontal and vertical directions relative to the high-fidelity model. The setting of the fluid solvers, other than the resolution, is the same for both models, with details presented in \cite{gong2021full}. The average wall time of high and low-fidelity simulations with 40 cores (Intel Xeon Gold 6154 CPU) are calibrated as 0.67 and 0.20 hours (considering adaptive time step size and parallel efficiency), leading to $c_h/c_l=3.36$ as the value we use in the sequential BED method. The time series of ship roll motion computed from the high and low fidelity models are shown in figure \ref{fig:cfd_results} for two examples with different input wave parameters, showing that the difference of the results from the two models (in general) increases with the group amplitude $A$. 

We use as the initial dataset 4 high-fidelity and 16 low-fidelity samples in the BF-O method, and 9 high-fidelity samples in the SF method (resulting in almost the same cost). Since the exact PDF in this case is unknown, we directly use the uncertainty level of the extreme-value PDF, i.e., $U$ as in \eqref{acq_L}, as an evaluation of the quality of results, which is plotted in figure \ref{fig:ship_results}(a) as a function of the computational cost measured by computation time in hours (only CFD times). It can be seen that the BF-O method results in a faster convergence of the extreme-value PDF compared to the SF method. This point is further illustrated in figure \ref{fig:ship_results}(b), which plots the PDFs from SF and BF-O method for the same computation time of 12 hours together with the upper and lower bounds (in terms of two standard deviations two-sided from the mean). It is clear that the PDF from the BF-O method is associated with appreciably lower uncertainty.

\section{Conclusion}
In this paper, we develop a multi-fidelity sequential Bayesian experimental design framework for efficient evaluation of the extreme response PDF. Our method leverages the multi-fidelity Gaussian process as a surrogate model, and a new acquisition function which allows the selection of the next-best sample in terms of both the location and fidelity level. We also construct an analytical formula for the acquisition function, which enables implementation of the method (e.g., gradient-based optimization) for high-dimensional problems. Our new method is first tested in a bi-fidelity context for a series of synthetic problems. With a broad range of low-fidelity accuracy level and computational cost, we show that the bi-fidelity method always outperforms the single-fidelity method, and that the BF-O method consistently shows advantage over the BF-F$n$ method, i.e., the bi-fidelity method with pre-defined fidelity hierarchy. We finally demonstrate the effectiveness of our BF-O method (relative to the single-fidelity method) in an engineering problem to evaluate the extreme ship roll PDF in irregular waves, where CFD of two resolutions serve as the high and low fidelity models. Further improvements on this framework may be achieved by considering reinforcement learning, which will be a topic of future study.

\section*{ACKNOWLEDGEMENT}

This research is supported by the Office of Naval Research
grant N00014-20-1-2096. We thank the program manager Dr. Woei-Min Lin for several helpful discussions
on the research. 
This work used the Extreme Science and Engineering Discovery Environment (XSEDE) through allocation TG-BCS190007.

\bibliographystyle{unsrt}
\bibliography{reference.bib}

\appendix

\section{Bi-fidelity Gaussian process}
The bi-fidelity Gaussian process conditioned on $\mathcal{D}=\{\{\mathcal{X}_1, \mathcal{X}_2\}, \{\mathcal{Y}_1, \mathcal{Y}_2\}\}$ can be considered as the counterpart of \eqref{mfgp} for $s=2$, with its mean and covariance formulated as 
\begin{align}
    \mathbb{E}(
    \begin{bmatrix} f_1(\mathbf{x}) \\ f_2(\mathbf{x}) \end{bmatrix}
   |\mathcal{D})  & = \rm{cov}(\begin{bmatrix} f_1(\mathbf{x}) \\ f_2(\mathbf{x}) \end{bmatrix}, \begin{bmatrix} \mathbf{Y}_1 \\ \mathbf{Y}_2 \end{bmatrix}) 
   {\rm{cov}}(\begin{bmatrix} \mathbf{Y}_1 \\ \mathbf{Y}_2 \end{bmatrix})^{-1} \begin{bmatrix} \mathcal{Y}_1 \\ \mathcal{Y}_2 \end{bmatrix},
\\  
    {\rm{cov}}\big(\begin{bmatrix} f_1(\mathbf{x}) \\ f_2(\mathbf{x}') \end{bmatrix} |\mathcal{D} \big) & = {\rm{cov}}\big(\begin{bmatrix} f_1(\mathbf{x}) \\ f_2(\mathbf{x}') \end{bmatrix}) - \rm{cov}( \begin{bmatrix} f_1(\mathbf{x}) \\ f_2(\mathbf{x}') \end{bmatrix}, \begin{bmatrix} \mathbf{Y}_1 \\ \mathbf{Y}_2 \end{bmatrix}) {\rm{cov}}(\begin{bmatrix} \mathbf{Y}_1 \\ \mathbf{Y}_2 \end{bmatrix})^{-1} \rm{cov}(\begin{bmatrix} \mathbf{Y}_1 \\ \mathbf{Y}_2 \end{bmatrix}, \begin{bmatrix} f_1(\mathbf{x}) \\ f_2(\mathbf{x}') \end{bmatrix}),
\end{align}
where
\begin{align}
    \rm{cov}( \begin{bmatrix} \mathbf{Y}_1 \\ \mathbf{Y}_2 \end{bmatrix})
    & = 
    \begin{bmatrix}
    k_1(\mathcal{X}_1,\mathcal{X}_1) + \gamma_1 \mathrm{I} & \rho_1 k_1(\mathcal{X}_1,\mathcal{X}_2) \\
    \rho_1 k_1(\mathcal{X}_2,\mathcal{X}_1) & \rho_1^2 k_1(\mathcal{X}_2,\mathcal{X}_2) + k_2(\mathcal{X}_2,\mathcal{X}_2) + \gamma_2 \mathrm{I}
    \end{bmatrix},
\\
    \rm{cov}( \begin{bmatrix} f_1(\mathbf{x}) \\ f_2(\mathbf{x}') \end{bmatrix}, \begin{bmatrix} \mathbf{Y}_1 \\ \mathbf{Y}_2 \end{bmatrix}) 
    & = 
    \begin{bmatrix} k_1(\mathbf{x}, \mathcal{X}_1) &  \rho_1 k_1(\mathbf{x}, \mathcal{X}_2) \\  \rho_1 k_1(\mathbf{x}', \mathcal{X}_1)  & \rho_1^2 k_1(\mathbf{x}', \mathcal{X}_2) + k_2(\mathbf{x}', \mathcal{X}_2) 
    \end{bmatrix}, 
\\
    {\rm{cov}}\big(\begin{bmatrix} f_1(\mathbf{x}) \\ f_2(\mathbf{x}') \end{bmatrix})
    & = 
    \begin{bmatrix} k_1(\mathbf{x}, \mathbf{x}) &  \rho_1 k_1(\mathbf{x}, \mathbf{x}') \\  \rho_1 k_1(\mathbf{x}', \mathbf{x})  & \rho_1^2 k_1(\mathbf{x}', \mathbf{x}') + k_2(\mathbf{x}', \mathbf{x}') 
    \end{bmatrix}. 
\end{align}

\section{Derivation of \eqref{tfgp_acq} using recursive update}

For the derivation of \eqref{tfgp_acq}, we consider the following Bayes' theorem:
\begin{align}
    p(f(\mathbf{x})|\mathcal{D}, \overline{y}_i(\tilde{\mathbf{x}})) = 
    \frac{p(f(\mathbf{x}), \overline{y}_i(\tilde{\mathbf{x}})|\mathcal{D})}{p( \overline{y}_i(\tilde{\mathbf{x}})|\mathcal{D})}.
\end{align}
where $f(\mathbf{x})|\mathcal{D}, \overline{y}_i(\tilde{\mathbf{x}})$ can be seen as the posterior of $f(\mathbf{x})|\mathcal{D}$ (as a prior) after adding one sample $\overline{y}_i(\tilde{\mathbf{x}})$. One can then get the mean and variance of $f(\mathbf{x})|\mathcal{D}, \overline{y}_i(\tilde{\mathbf{x}})$ using the standard conditional Gaussian formula:
\begin{align}
    \mathbb{E}(f(\mathbf{x})|\mathcal{D}, \overline{y}_i(\tilde{\mathbf{x}}))  & = \mathbb{E}(f(\mathbf{x})|\mathcal{D}) +  \frac{{\rm{cov}}(f(\mathbf{x}), f_i(\tilde{\mathbf{x}})| \mathcal{D})(\overline{y}_i(\tilde{\mathbf{x}})-\mathbb{E}(f_i(\tilde{\mathbf{x}})|\mathcal{D}))}{{\rm{var}}(y_i(\tilde{\mathbf{x}})|\mathcal{D})},
\\
    {\rm{var}}\big(f(\mathbf{x})|\mathcal{D}, \overline{y}_i(\tilde{\mathbf{x}})\big) 
    & = {\rm{var}}\big(f(\mathbf{x})|\mathcal{D}) -
    \frac{{\rm{cov}}(f(\mathbf{x}),
    f_i(\tilde{\mathbf{x}})| \mathcal{D})^2}
    {{\rm{var}}(y_i(\tilde{\mathbf{x}})|\mathcal{D})}.
\label{recur_cov}
\end{align}

The formula in \eqref{tfgp_acq} is a direct result of \eqref{recur_cov}.

\section{Example of a more informative low-fidelity sample}
We show a special case in Theorem 1 where a high-fidelity sample is less informative than a low-fidelity sample in a bi-fidelity model ($s=2$). 

\textbf{Theorem 1:} Assume $\rho_1=1$ and $\gamma_i=0$ with $i=1,2$ (see \eqref{noise} and \eqref{AR}). The benefits of adding a high-fidelity sample $B(2, \tilde{\mathbf{x}})$ is smaller than  $B(1, \tilde{\mathbf{x}})$ for $\|\Lambda_2\| \rightarrow 0$.

\textit{Proof:} 
Based on \eqref{tfgp_acq}, \eqref{cov}, \eqref{tf_cov_prior}, we can compute the benefits as
\begin{align}
    B(1, \tilde{\mathbf{x}})  & = Q(\mathcal{D}) - Q(\mathcal{D}, 1, \tilde{\mathbf{x}})
\nonumber \\
    & = \frac{1}{{\rm{var}}(y_1(\tilde{\mathbf{x}})|\mathcal{D})}\int {\rm{cov}}^2(f_{}(\mathbf{x}),f_1(\tilde{\mathbf{x}})|\mathcal{D}) w(\mathbf{x}) \mathrm{d} \mathbf{x}
\nonumber \\
    & = \int \frac{(\mathrm{cov}(f(\mathbf{x}), f_1(\tilde{\mathbf{x}}))-\mathrm{cov}(f(\mathbf{x}), \mathbf{Y})\mathrm{cov}(\mathbf{Y})^{-1} \mathrm{cov}(\mathbf{Y},f_1(\tilde{\mathbf{x}})))^2}
    {\mathrm{cov}(f_1(\tilde{\mathbf{x}})) - \mathrm{cov}(f_1(\tilde{\mathbf{x}}), \mathbf{Y})\mathrm{cov}(\mathbf{Y})^{-1} \mathrm{cov}(\mathbf{Y},f_1(\tilde{\mathbf{x}}))} 
    w(\mathbf{x}) \mathrm{d} \mathbf{x}
\nonumber \\
    & = \int \frac{(\splitfrac{k_1(\mathbf{x}, \tilde{\mathbf{x}})-[k_1(\mathbf{x}, \mathcal{X}_1), k_1(\mathbf{x}, \mathcal{X}_2)+k_2(\mathbf{x}, \mathcal{X}_2)]}{ \mathrm{cov}(\mathbf{Y})^{-1} [k_1(\tilde{\mathbf{x}}, \mathcal{X}_1), k_1(\tilde{\mathbf{x}}, \mathcal{X}_2)]^T})^2}
    {\splitfrac{k_1(\tilde{\mathbf{x}}, \tilde{\mathbf{x}}) - [k_1(\tilde{\mathbf{x}}, \mathcal{X}_1), k_1(\tilde{\mathbf{x}}, \mathcal{X}_2)]} {\mathrm{cov}(\mathbf{Y})^{-1} [k_1(\tilde{\mathbf{x}}, \mathcal{X}_1), k_1(\tilde{\mathbf{x}}, \mathcal{X}_2)]^T}}
    w(\mathbf{x}) \mathrm{d} \mathbf{x},
\\
    B(2, \tilde{\mathbf{x}})  & = Q(\mathcal{D}) - Q(\mathcal{D}, 2, \tilde{\mathbf{x}})
\nonumber \\
    & = \frac{1}{{\rm{var}}(y_2(\tilde{\mathbf{x}})|\mathcal{D})}\int {\rm{cov}}^2(f_{}(\mathbf{x}),f_2(\tilde{\mathbf{x}})|\mathcal{D}) w(\mathbf{x}) \mathrm{d} \mathbf{x}
\nonumber \\
    & = \int \frac{(\mathrm{cov}(f(\mathbf{x}), f_2(\tilde{\mathbf{x}}))-\mathrm{cov}(f(\mathbf{x}), \mathbf{Y})\mathrm{cov}(\mathbf{Y})^{-1} \mathrm{cov}(\mathbf{Y},f_2(\tilde{\mathbf{x}})))^2}
    {\mathrm{cov}(f_2(\tilde{\mathbf{x}})) - \mathrm{cov}(f_2(\tilde{\mathbf{x}}), \mathbf{Y})\mathrm{cov}(\mathbf{Y})^{-1} \mathrm{cov}(\mathbf{Y},f_2(\tilde{\mathbf{x}}))} 
    w(\mathbf{x}) \mathrm{d} \mathbf{x}
\nonumber \\
    & = \int \frac{(\splitfrac{k_1(\mathbf{x}, \tilde{\mathbf{x}}) + k_2(\mathbf{x}, \tilde{\mathbf{x}}) -[k_1(\mathbf{x}, \mathcal{X}_1), k_1(\mathbf{x}, \mathcal{X}_2)+k_2(\mathbf{x}, \mathcal{X}_2)]}{\mathrm{cov}(\mathbf{Y})^{-1} [k_1(\tilde{\mathbf{x}}, \mathcal{X}_1), k_1(\tilde{\mathbf{x}}, \mathcal{X}_2) + k_2(\tilde{\mathbf{x}}, \mathcal{X}_2)]^T})^2}
    {\splitfrac{k_1(\tilde{\mathbf{x}}, \tilde{\mathbf{x}}) + k_2(\tilde{\mathbf{x}}, \tilde{\mathbf{x}}) - [k_1(\tilde{\mathbf{x}}, \mathcal{X}_1), k_1(\tilde{\mathbf{x}}, \mathcal{X}_2) + k_2(\tilde{\mathbf{x}}, \mathcal{X}_2)] }{\mathrm{cov}(\mathbf{Y})^{-1} [k_1(\tilde{\mathbf{x}}, \mathcal{X}_1), k_1(\tilde{\mathbf{x}}, \mathcal{X}_2) + k_2(\tilde{\mathbf{x}}, \mathcal{X}_2)]^T}} 
    w(\mathbf{x}) \mathrm{d} \mathbf{x}.
\label{tfgp_acq_extreme}
\end{align}
If $\|\Lambda_2\| \rightarrow 0$, we have $k_2(\mathbf{x}, \mathbf{x}'; \Lambda_2) \rightarrow 0 $ for $\mathbf{x} \neq \mathbf{x}'$ (see \eqref{RBF}). Thus 
\begin{align}
    \lim_{\|\Lambda_2\| \to 0} B(1, \tilde{\mathbf{x}}) & = \int \frac{(k_1(\mathbf{x}, \tilde{\mathbf{x}})- k_1(\mathbf{x}, \mathcal{X}) \mathrm{cov}(\mathbf{Y})^{-1} k_1(\tilde{\mathbf{x}}, \mathcal{X})^T)^2}
    {k_1(\tilde{\mathbf{x}}, \tilde{\mathbf{x}}) - k_1(\tilde{\mathbf{x}}, \mathcal{X})\mathrm{cov}(\mathbf{Y})^{-1} k_1(\tilde{\mathbf{x}}, \mathcal{X})^T}
    w(\mathbf{x}) \mathrm{d} \mathbf{x},
\\ 
    \lim_{\|\Lambda_2\| \to 0} B(2, \tilde{\mathbf{x}}) 
    & = \int \frac{(k_1(\mathbf{x}, \tilde{\mathbf{x}})- k_1(\mathbf{x}, \mathcal{X}) \mathrm{cov}(\mathbf{Y})^{-1} k_1(\tilde{\mathbf{x}}, \mathcal{X})^T)^2}
    {k_1(\tilde{\mathbf{x}}, \tilde{\mathbf{x}}) + k_2(\tilde{\mathbf{x}}, \tilde{\mathbf{x}}) - k_1(\tilde{\mathbf{x}}, \mathcal{X})\mathrm{cov}(\mathbf{Y})^{-1} k_1(\tilde{\mathbf{x}}, \mathcal{X})^T}
    w(\mathbf{x}) \mathrm{d} \mathbf{x},
\end{align}
and
\begin{equation}
    \lim_{\|\Lambda_2\| \to 0} B(1, \tilde{\mathbf{x}}) - B(2, \tilde{\mathbf{x}}) > 0,
\end{equation}
which complete the proof.

\section{Analytic computation of the acquisition}
\sloppy We first present the formula of $G_t(f_i(\mathbf{x}_1), f_j(\mathbf{x}_2))$ in \eqref{G}, which is the key part towards an analytical acquisition and its derivative. Substituting the covariance function \eqref{tf_cov_prior} to \eqref{G}, one obtains
\begin{align}
    & G_t(f_i(\mathbf{x}_1), f_j(\mathbf{x}_2))
\nonumber \\
    = &  \int {\rm{cov}}\big(f_i(\mathbf{x}_1), f_{}(\mathbf{x})\big) {\rm{cov}}\big(f_{}(\mathbf{x}), f_j(\mathbf{x}_2))\big) \mathcal{N}(\mathbf{x};\mathbf{\mu}_t, \Sigma_t) \mathrm{d} \mathbf{x}
\nonumber \\
    = &  \int \big(\sum_{l=1}^{i} \pi_{isl} k_l(\mathbf{x}_1, \mathbf{x})\big) \big(\sum_{r=1}^{j} \pi_{jsr} k_r(\mathbf{x}, \mathbf{x}_2) \big) \mathcal{N}(\mathbf{x};\mathbf{\mu}_t, \Sigma_t) \mathrm{d} \mathbf{x}
\nonumber \\ 
    = & \sum_{l=1}^{i} \sum_{r=1}^{j}  \pi_{isl}  \pi_{jsr} \mathcal{I}_{l,r,t}(\mathbf{x}_1, \mathbf{x}_2),
\label{G_t}
\end{align}
where
\begin{subequations}
\begin{align}
    \mathcal{I}_{l,r,t}(\mathbf{x}_1, \mathbf{x}_2) = &
    \int k_l(\mathbf{x}_1, \mathbf{x}; \Lambda_l)  k_r(\mathbf{x}_2, \mathbf{x}; \Lambda_r) 
    \mathcal{N}(\mathbf{x};\mathbf{\mu}_t, \Sigma_t) \mathrm{d} \mathbf{x} 
\label{I1} \\
     = & |\Sigma_t \mathrm{M}^{-1} + \mathrm{I}|^{-\frac{1}{2}}  k_l(\mathbf{x}_1, \mathbf{x}_2;\Lambda_l+\Lambda_r)
    k_r(\mathbf{m},\mathbf{\mu}_t; \Sigma_t+ \mathrm{M}),
\label{I3}
\end{align}
\end{subequations}
with
\begin{align}
    \mathbf{m} = & \Lambda_r(\Lambda_l+\Lambda_r)^{-1}\mathbf{x}_1 + \Lambda_l(\Lambda_l+\Lambda_r)^{-1}\mathbf{x}_2,
\\
    \mathrm{M} = & \Lambda_l \Lambda_r (\Lambda_l + \Lambda_r)^{-1}.
\end{align}
We note that from \eqref{I1} to \eqref{I3} one needs to use the formulae for transferring kernels to Gaussian functions as well as the multiplication of Gaussian functions \cite{rasmussen2003gaussian}:
\begin{align}
    k(\mathbf{x}, \mathbf{x}_1;\Lambda, \tau) & = \tau^2 (2\pi)^{d/2} |\Lambda|^{1/2} \mathcal{N}(\mathbf{x};\mathbf{x}_1, \Lambda), 
\label{kk} \\ 
    \mathcal{N}(\mathbf{x}; \mathbf{x}_1, \Sigma_1) \mathcal{N}(\mathbf{x}; \mathbf{x}_2, \Sigma_2) & = \mathcal{N}(\mathbf{x}_1; \mathbf{x}_2, \Sigma_1 + \Sigma_2) 
    \mathcal{N}(\mathbf{x}; \mathrm{C}(\Sigma_1^{-1}\mathbf{x_1}+\Sigma_2^{-1}\mathbf{x_2}), \mathrm{C}),
\label{kN} 
\end{align}
with $\mathrm{C} = (\Sigma_1^{-1} + \Sigma_2^{-1})^{-1}$.

Specifically, the detailed derivation of \eqref{I3} using \eqref{kk} and \eqref{kN} is shown below:
\begin{subequations}
\begin{align}
    \mathcal{I}_{l,r,t}(\mathbf{x}_1, \mathbf{x}_2) = &
    \int k_l(\mathbf{x}_1, \mathbf{x}; \Lambda_l, \tau_l)  k_r(\mathbf{x}_2, \mathbf{x}; \Lambda_r, \tau_r) 
    \mathcal{N}(\mathbf{x};\mathbf{\mu}_t, \Sigma_t) \mathrm{d} \mathbf{x} 
\\
    = &  \tau_l^2 \tau_r^2 (2\pi)^d |\Lambda_l|^{1/2}|\Lambda_r|^{1/2} \int
    \mathcal{N}(\mathbf{x}; \mathbf{x}_1, \Lambda_l) \mathcal{N}(\mathbf{x}; \mathbf{x}_2, \Lambda_r) \mathcal{N}(\mathbf{x};\mathbf{\mu}_t, \Sigma_t) \mathrm{d} \mathbf{x} 
\\
    = &  \tau_l^2 \tau_r^2 (2\pi)^d |\Lambda_l|^{1/2}|\Lambda_r|^{1/2}
    \mathcal{N}(\mathbf{x}_1; \mathbf{x}_2, \Lambda_l + \Lambda_r) \int \mathcal{N}(\mathbf{x}; \mathbf{m}, \mathrm{M}) \mathcal{N}(\mathbf{x};\mathbf{\mu}_t, \Sigma_t) \mathrm{d} \mathbf{x}
\\
    = & \tau_l^2 \tau_r^2 (2\pi)^d |\Lambda_l|^{1/2}|\Lambda_r|^{1/2}
    \mathcal{N}(\mathbf{x}_1; \mathbf{x}_2, \Lambda_l + \Lambda_r)  \mathcal{N}(\mathbf{m};\mathbf{\mu}_t, \mathrm{M} + \Sigma_t) 
\\
     = & |\Sigma_t \mathrm{M}^{-1} + \mathrm{I}|^{-\frac{1}{2}}  k_l(\mathbf{x}_1, \mathbf{x}_2;\Lambda_l+\Lambda_r)
    k_r(\mathbf{m},\mathbf{\mu}_t; \Sigma_t + \mathrm{M}).
\end{align}
\end{subequations}
With the analytical formula for $G_t(f_i(\mathbf{x}_1), f_j(\mathbf{x}_2))$ available in \eqref{G_t}, the benefit $B(i,\tilde{\mathbf{x}})$ in \eqref{tfgp_acq2} can be derived accordingly. To further obtain the analytical derivative of $B(i,\tilde{\mathbf{x}})$, we rewrite $B(i,\tilde{\mathbf{x}}) = T(i,\tilde{\mathbf{x}})/ {\rm{var}}(y_i(\tilde{\mathbf{x}})|\mathcal{D})$ (comparing to \eqref{tfgp_acq}), and derive the derivative of $T$ as
\begin{align}
        \frac{\partial T(i, \tilde{\mathbf{x}})}{\partial \tilde{\mathbf{x}}} =  &\frac{\partial \mathcal{K}(f_i(\tilde{\mathbf{x}}), f_i(\tilde{\mathbf{x}})) }{ \partial \tilde{\mathbf{x}}} + 2 \frac{\partial {\rm{cov}}(\mathbf{Y}, f_i(\tilde{\mathbf{x}}))}{\partial \tilde{\mathbf{x}}}  {\rm{cov}}(\mathbf{Y})^{-1}  \mathcal{K}(\mathbf{Y},\mathbf{Y})  {\rm{cov}}(\mathbf{Y})^{-1}  {\rm{cov}}(\mathbf{Y},f_i(\tilde{\mathbf{x}}))
\nonumber \\
        - 2 &\frac{\partial {\rm{cov}}(\mathbf{Y},f_i(\tilde{\mathbf{x}}))}{\partial \tilde{\mathbf{x}}}{\rm{cov}}(\mathbf{Y})^{-1} \mathcal{K}(\mathbf{Y}, f_i(\tilde{\mathbf{x}})) 
        - 2 \frac{\partial \mathcal{K}(\mathbf{Y}, f_i(\tilde{\mathbf{x}}))}{\partial \tilde{\mathbf{x}}} {\rm{cov}}(\mathbf{Y})^{-1} {\rm{cov}}(\mathbf{Y}, f_i(\tilde{\mathbf{x}})).
\label{T_deriv}
\end{align}
The analytical computation of \eqref{T_deriv} requires the formula for the derivative of the covariance function and $\mathcal{K}$. For the former, we have:
\begin{align}
    \frac{\partial {\rm{cov}}(f_i(\mathbf{x}_1), f_j(\mathbf{x}_2))}{\partial \mathbf{x}_1} = \sum_{l=1}^{\min(i,j)} \pi_{ijl} \frac{\partial k_l(\mathbf{x}_1, \mathbf{x}_2)}{\partial \mathbf{x}_1},
\end{align}
with 
\begin{equation}
    \frac{\partial k(\mathbf{x}_1, \mathbf{x}_2;\Lambda)}{\partial \mathbf{x}_1} = k(\mathbf{x}_1, \mathbf{x}_2;\Lambda)\Lambda^{-1} (\mathbf{x}_2- \mathbf{x}_1) .
\label{kernel_deriv}
\end{equation}
For the latter, we have:
\begin{align}
    \frac{\partial \mathcal{K}(f_i(\mathbf{x}_1), f_i(\mathbf{x}_1)) }{ \partial \mathbf{x}_1} 
    & = \sum_{t=1}^{n_{GMM}} \alpha_t \frac{\partial G_{t}(f_i(\mathbf{x}_1), f_i(\mathbf{x}_1))}{\partial \mathbf{x}_1}
\nonumber \\
    & = \sum_{t=1}^{n_{GMM}} \alpha_t \sum_{l=1}^{i} \sum_{r=1}^{i}  \pi_{isl}  \pi_{isr} \frac{\partial \mathcal{I}_{l,r,t}(\mathbf{x}_1, \mathbf{x}_1)}{\partial \mathbf{x}_1},
\label{K_d_1} \\
     \frac{\partial \mathcal{K}(f_i(\mathbf{x}_1), f_j(\mathbf{x}_2)) }{ \partial \mathbf{x}_1} 
     & = \sum_{t=1}^{n_{GMM}} \alpha_t  \frac{\partial G_{t}(f_i(\mathbf{x}_1), f_j(\mathbf{x}_2))}{\partial \mathbf{x}_1}
\nonumber \\ 
     & = \sum_{t=1}^{n_{GMM}} \alpha_t \sum_{l=1}^{i} \sum_{r=1}^{j}  \pi_{isl}  \pi_{jsr} \frac{\partial \mathcal{I}_{l,r,t}(\mathbf{x}_1, \mathbf{x}_2)}{\partial \mathbf{x}_1}.
\label{K_d_2}
\end{align}
Finally, \eqref{K_d_1} and \eqref{K_d_2} require the derivatives of $\mathcal{I}_{i,j,t}$, which can be obtained by combing \eqref{I3} and \eqref{kernel_deriv}:
\begin{align}
    \frac{\partial \mathcal{I}_{l,r,t}(\mathbf{x}_1, \mathbf{x}_1)}{\partial \mathbf{x}_1} = &  \mathcal{I}_{l,r,t}(\mathbf{x}_1, \mathbf{x}_1) 
    (\Sigma_t + \mathrm{M})^{-1}(\mathbf{\mu}_t - \mathbf{x}_1),
\\
        \frac{\partial \mathcal{I}_{l,r,t}(\mathbf{x}_1, \mathbf{x}_2)}{\partial \mathbf{x}_1} = &  \mathcal{I}_{l,r,t}(\mathbf{x}_1, \mathbf{x}_2) 
    \Big((\Lambda_l + \Lambda_r)^{-1}(\mathbf{x}_2-\mathbf{x}_1)
    + \Lambda_r(\Lambda_l+\Lambda_r)^{-1}(\Sigma_t + \mathrm{M})^{-1}(\mathbf{\mu}_t - \mathbf{m})
    \Big).
\end{align}
\end{document}